\documentclass{iopjournal}
\usepackage{algorithm}
\usepackage{color}
\usepackage{algorithmic}
\usepackage{amsmath}
\usepackage{braket}
\usepackage{multirow}
\usepackage{amssymb}
\usepackage{amsthm}
\usepackage{bm}
\usepackage{thmtools}

\usepackage{array}
\usepackage{url}
\usepackage{hyperref}
\hypersetup{
    colorlinks=true,
    linkcolor=blue,
    filecolor=magenta,      
    urlcolor=cyan,
    citecolor=red
}
\usepackage{graphicx}
\usepackage{subcaption}
\usepackage{bibentry}

\newtheorem*{theorem*}{Theorem}

\newcommand{\R}{\mathbb{R}}

\begin{document}

\articletype{Paper} 

\title{Heralded enhancement in quantum state discrimination}

\author{Qipeng Qian$^{1,*}$\orcid{0009-0003-2407-1292}, Christos N. Gagatsos$^{1,2,3}$\orcid{0000-0000-0000-0000}}

\affil{$^1$Program in Applied Mathematics, The University of Arizona, Tucson, Arizona 85721, USA}

\affil{$^2$Department of Electrical and Computer Engineering, The University of Arizona, Tucson, Arizona, 85721, USA}

\affil{$^3$Wyant College of Optical Sciences, The University of Arizona, Tucson, Arizona, 85721, USA}

\affil{$^*$Author to whom any correspondence should be addressed.}

\email{qqian@arizona.edu}


\begin{abstract}
The discrimination of quantum states is a central problem in quantum information science and technology. Meanwhile, partial post-selection has emerged as a valuable tool for quantum state engineering. In this work, we bring these two areas together and ask whether partial measurements can enhance the discrimination performance between two unknown and non-orthogonal pure states. Our framework is general: the two unknown states interact with the same environment—set in a pure state—via an arbitrary unitary transformation. A measurement is then performed on one of the output modes (i.e. a partial measurement), modeled by an arbitrary positive operator-valued measure (POVM). We then allow classical communication to inform the unmeasured mode of the outcome of the partial measurement, which is subsequently measured by a POVM that is optimal in the sense that the discrimination probability of error is minimized. The two POVMs act locally and classical information is exchanged between the two modes, representing a single-round (feed-forward) form of local operations with classical communication. Under these considerations, we first show that, as expected, the minimum error probability, averaged over all possible conditional states, cannot be reduced below the minimum error probability of discriminating the original input states.  Then, we devise a generic setup produces specific examples where the conditional discrimination can achieve strictly lower error probabilities than the original optimal measurement, illustrating that while post-selection does not improve the average performance, it can enable better discrimination in certain post-selected ensembles. 
\end{abstract}

\section{Introduction}\label{sec:intro}

Quantum state discrimination is a central problem in quantum information theory, concerned with identifying an unknown quantum state, from a known set of states, with minimum error probability \cite{Helstrom1976,YKL1975,Holevo1973}. Because non-orthogonal quantum states cannot be perfectly distinguished, this task is inherently probabilistic and requires optimizing over measurement strategies to minimize the error. While the discrimination of quantum states is fundamentally important on its own, it also underpins a wide range of quantum applications. In quantum communication, efficient discrimination enables reliable decoding of quantum-encoded messages over noisy channels \cite{sidhu2023linear,Rossinoisy}. In quantum cryptography, the security often depends on the inability of an adversary to perfectly distinguish between non-orthogonal quantum states \cite{Bennettcryptography}. In quantum metrology and sensing, discrimination tasks assist in estimating physical parameters, such as phase shifts or displacements \cite{Tan2008Illumination}. Alongside the applications, many theoretical advances have been reported, such as closed-form formulas for the Chernoff exponent for the case of Gaussian states \cite{Pirandola2008Computable}.

A widely studied figure of merit in quantum discrimination is the minimum error probability, where the goal is to minimize the average probability of guessing the wrong state. For the binary case of two pure states $|\psi_1\rangle$ and $|\psi_2\rangle$ presented with prior probabilities $1-q$ and $q$ respectively, the probability of error is defined as,
\begin{eqnarray}
    P_{err}=(1-q)\text{Tr}[\hat{E}_2|\psi_1\rangle\langle\psi_1|]+q\text{Tr}[\hat{E}_1|\psi_2\rangle\langle\psi_2|],
    \label{eq:P_err}
\end{eqnarray}
where $\{\hat{E}_1,\hat{E}_2\}$ is any positive operator-valued measure (POVM) used for discrimination. The minimum probability of error achievable by any POVM is given by the Helstrom bound \cite{Helstrom1976}
\begin{eqnarray}
    P_{err}\geq\frac{1}{2}-\frac{\|\hat{\rho}\|_1}{2}=:P_{me},
    \label{eq:me}
\end{eqnarray}
where $\hat{\rho}=(1-q)|\psi_1\rangle\langle\psi_1|-q|\psi_2\rangle\langle\psi_2|$ and $\|.\|_1$ is the trace norm. The equality in Eq. \eqref{eq:me}
is achieved by the Helstrom
measurement and represents the ultimate theoretical limit for any measurement aiming to distinguish two quantum states under the symmetric minimum-error criterion \cite{RegulaPostselected}.

In contrast to quantum discrimination protocols that rely on a single, global (joint) measurement, post-selection introduces an intermediate measurement step that conditions the subsequent processing of the system. Specifically, post-selection involves first performing measurements on part of the quantum system, and then, depending on the outcome, choosing whether and how to proceed with further operations pertaining, for example, to state discrimination. 
In many platform-level protocols, performance is evaluated conditional on a heralded success event  (in the sense that any successful events are announced by triggering an accessible to the user alarm e.g. a photon number detector), rather than averaged over all branches: e.g., heralded entanglement generation or swapping in quantum-repeater architectures \cite{Duan2001DLCZ} and probabilistic, measurement-induced gates in linear-optical implementations, where one proceeds only upon specific detector patterns \cite{Knill2001KLM,Scheel2006FeedForward}. This motivates studying branch-wise (post-selected) discrimination accuracy and its trade-off with success probability, even when the ensemble-averaged error cannot be improved \cite{BaeKwek2015Review}. A canonical illustration within state discrimination itself is unambiguous state discrimination (USD), which can be viewed as an extreme form of post-selection: one post-selects on conclusive outcomes and discards the inconclusive branch \cite{CHEFLES1998339, Clarke2001Experimental}. Such post-selection strategies are particularly prominent in quantum optics and continuous-variable protocols, where conditional operations (e.g., photon subtraction or partial homodyne detection) have been used for quantum state engineering \cite{Lance2006Quantum-state,Su2019,Eaton2019,Gagatsos2019} and generating of photonic cluster states \cite{Guichard_2026}, teleportation \cite{Kok2000Postselected}, quantum computing \cite{Lloyd2011Closed,Eaton2022}, just to name a few. Beyond post-selection, enhanced discrimination can also be pursued via entanglement-assisted schemes or collective (joint) measurements across multiple copies \cite{PianiWatrous2009PRL}. In contrast, our framework explicitly couples the system to an environment (or auxiliary) mode and performs an intermediate partial measurement on that mode. In this way, the auxiliary mode serves as an operational ancilla that enables outcome-conditioned strategies, allowing us to target conditional performance improvements without invoking multi-copy joint measurements or preshared entanglement \cite{Kok2007RMP}. Ancilla-assisted implementations of nontrivial discrimination measurements are in fact widespread in optical platforms, where additional modes, auxiliary coherent fields, or extra photonic degrees of freedom are leveraged to realize (near-)optimal receivers and measurement statistics (see, e.g., \cite{Sol2017Experimental,Sidhu_2021,Laneve_2022}). 

While post-selection has found various applications in quantum information processing, how well it performs in general does not have a unique answer. For instance, in the context of quantum metrology, it has been rigorously shown that post-selected probabilistic schemes cannot surpass the quantum limits for single-parameter estimation—neither on average nor asymptotically in the number of trials—when the performance is measured by the mean-square error \cite{Combespostselected}, unless if the environment is continuously monitored \cite{Albarelli2018}. Related questions about the operational role of post-selection and ancillary systems have also been explored in single-shot discrimination of quantum measurements, where outcome-conditioned strategies and explicit inconclusive branches are analyzed \cite{SedlakZiman2014}. Complementing these theoretical insights, a number of experiments implement optimal (or near-optimal) discrimination measurements by coupling to ancillary modes or leveraging additional degrees of freedom, see e.g. \cite{Sol2017Experimental,Sidhu_2021,Laneve_2022}. This raises a natural question as to whether similar limitations apply in other quantum information tasks, such as state discrimination which is discussed in this paper. 

In this work, we investigate whether post-selection can enhance—or possibly degrade—the ability to distinguish between quantum states. Unlike USD, which requires error-free conclusive outcomes by allowing inconclusive results, our approach focuses only on conclusive discrimination and does not require zero error. Rather, it can be viewed as interpolating between MED (minimum error discrimination) and USD: it seeks to reduce the error probability as in MED, while approaching USD in the extreme case where the conditional error of a retained branch vanishes at the cost of a reduced success probability. We also clarify when post-selection can matter: the measurement outcome on one subsystem can affect the discrimination task on the other only if the joint unitary $U$ generates nontrivial correlations between them. As shown in Example 2 and Appendix C, no such effect is observed when no entanglement is created.

This paper is organized as follows: In Section \ref{sec:model}, we introduce the post-selection and state-engineering-based scheme. We first introduce the setup of our model and the state discrimination procedure. Then we point out the consequent expected probability of error that serves as the performance figure of merit. In Section \ref{subsec:no go proof}, we prove the post-selection and state-engineering-based scheme considered in this work will only degrade the state discrimination ability under the introduced performance criterion. In Section \ref{subsec: Example}, we give a concrete example to explicitly demonstrate our general result which also shows that one can perform better than the minimum error probability in a probabilistic yet heralded way. In Section \ref{sec:conclusion}, we briefly summarize our results and discuss the intuition behind our findings.

\section{The model and mathematical setting}\label{sec:model}
In this section, we specify the post-selection model that we consider and the corresponding mathematical setups. We focus on the task of discriminating between two pure quantum states, $|\psi_1\rangle$ and $|\psi_2\rangle$, presented with prior probabilities $1-q$ and $q$ respectively. In contrast to previous works on post-selection methods \cite{combes2015cost}, our model incorporates an interaction with a given environment, represented by a pure state $|e\rangle$. The unknown input state $|\psi_i\rangle,\ i=1,2$, and the environment state $|e\rangle$ are jointly processed through an arbitrary unitary transformation, denoted by $\hat{U}$, which in general entangles input and environment.

\begin{figure}[h]
    \centering
    \includegraphics[width=0.8\textwidth]{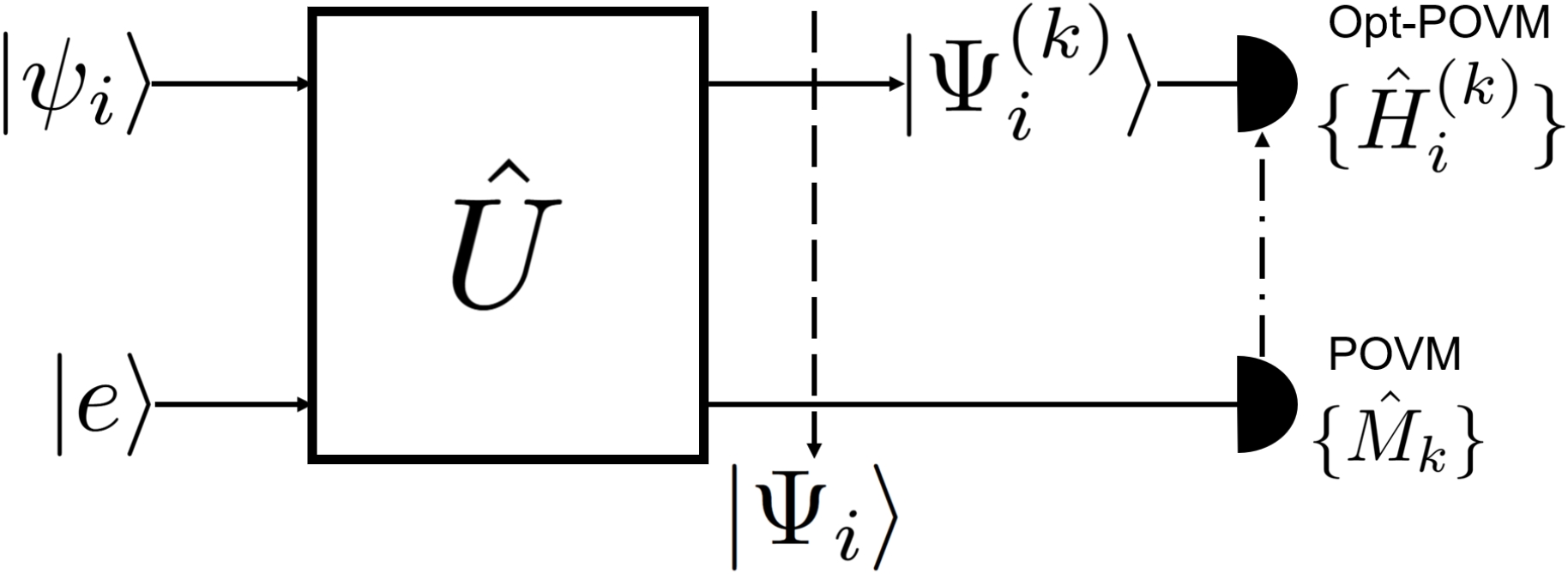}
    \caption{Post-selection model: The input state $|\psi_i\rangle$ (where $i=1,2$) interacts with the environment $|e\rangle$ through a unitary $\hat{U}$. The resulting state $|\Psi_i\rangle$ undergoes partial measurement described by a POVM $\{\hat{M}_k\}$, leaving the undetected part of the system in the conditional state $|\Psi_i^{(k)}\rangle$ which will then be discriminated by POVM $\{\hat{H}_i^{(k)}\}$ which represents the Helstrom measurement and depends on the outcome $k$.}
    \label{fig: model}
\end{figure}

Following this transformation, we perform a partial measurement on the second mode using a POVM $\{\hat{M}_k\}$. Based on the measurement outcome $k$, the resulting conditional state on the first mode—denoted by $|\Psi_i^{(k)}\rangle$—is obtained. In our model we allow the exchange of classical information between the two modes. That is, we then apply on the upper output mode a POVM $\{\hat{H}_i^{(k)}\}$, which depends on the outcome of the previous measurement on the lower mode, to distinguish between the possible states $|\Psi_1^{(k)}\rangle$ and $|\Psi_2^{(k)}\rangle$, thereby inferring whether the original unknown input state was $|\psi_1\rangle$ or $|\psi_2\rangle$ respectively. In this work, the POVM on the upper output mode is chosen to be the Helstrom measurement, i.e., the measurement that discriminates $|\Psi_1^{(k)}\rangle$ and $|\Psi_2^{(k)}\rangle$ with minimum probability of error.  Our method pertains to a single-round adaptive (feed-forward) local operations and classical communication (LOCC) protocol. The overall structure of the model and the discrimination procedure are shown in Fig. \ref{fig: model}. 

Previous studies on local operations and classical communication (LOCC) mainly focus on multi-copy or asymptotic discrimination scenarios.
For example, \cite{Acin2005} investigated the performance of local (including adaptive) measurements in multi-copy state discrimination, and \cite{Cals2010} demonstrated a persistent error-rate gap between LOCC and collective measurements as the number of copies increases.
More recent works, such as \cite{Cohen2023, Heo2024, Conlon2025}, further developed bounds and necessary and/or sufficient conditions for asymptotic LOCC discrimination and analyzed when fully entangling collective measurements are required to saturate the Helstrom bound.

In contrast, our work considers a single-copy scenario with only one round of classical communication, where the feed-forward structure determines the second local measurement.
This single-round, one-way adaptive framework fundamentally differs from the multi-round adaptive LOCC protocols studied in Refs.~\cite{Acin2005,Cals2010,Cohen2023,Heo2024,Conlon2025}.
Moreover, it enables a post-selection-based decomposition of discrimination performance: certain conditional subensembles can outperform the Helstrom bound, even though the overall average remains bounded by it.
Such a feed-forward LOCC structure is not only operationally simpler and experimentally more feasible than multi-round adaptive schemes but also provides new insight into how discrimination advantages can emerge conditionally within a single-copy setting. 

We begin by computing the probability of discrimination error for the conditional state remaining on the first mode, given that the measurement outcome on the second mode is $k$. This conditional error probability, analogous to Eq. (\ref{eq:P_err}), is denoted by $P_{\mathrm{err}}^{(k)}$. Then, following Eq. (\ref{eq:me}), we have 
\begin{eqnarray}
    P_{err}^{(k)}&=&(1-q)\text{Tr}[\hat{E}_2^{(k)}|\Psi_1^{(k)}\rangle\langle\Psi_1^{(k)}|]\nonumber\\
    &&+q\text{Tr}[\hat{E}_1^{(k)}|\Psi_2^{(k)}\rangle\langle\Psi_2^{(k)}|]\nonumber\\
    &\geq&\frac{1}{2}-\frac{\|\hat{\rho}^{(k)}\|_1}{2}\nonumber\\
    &=:&P_{me}^{(k)},
    \label{eq:P_err^(k)}
\end{eqnarray}
where $\{\hat{E}^{(k)}_1,\hat{E}^{(k)}_2\}$ is any POVM used for discrimination of the conditional states and 
\begin{eqnarray}
    \hat{\rho}^{(k)}=P'(1|k)|\Psi_1^{(k)}\rangle\langle\Psi_1^{(k)}|-P'(2|k)|\Psi_2^{(k)}\rangle\langle\Psi_2^{(k)}|. 
    \label{eq:rho^(k)}
\end{eqnarray}
Here, we use $P'(i|k)$ to represent the prior probability of $|\Psi_i^{(k)}\rangle$. To write $\hat{\rho}^{(k)}$ explicitly, we first have,
\begin{eqnarray}
    |\Psi_i^{(k)}\rangle = \frac{(\hat{I} \otimes \hat{M}_k^{1/2}) \hat{U}|\psi_i,e\rangle}{\sqrt{P(k|i)}}, 
    \label{eq:Psi_i^(k)}
\end{eqnarray}
where
\begin{eqnarray}
    P(k|i)=\langle\psi_i,e|\hat{U}^{\dag}(\hat{I}\otimes \hat{M}_k)\hat{U}|\psi_i,e\rangle 
    \label{eq:P(k|i)}
\end{eqnarray}
denotes the probability of getting the measurement outcome $k$ given the input state is $|\psi_i\rangle$. Then, we use Bayes' theorem to calculate the post-selected prior probability $P'(i|k)$ as follows, 
\begin{eqnarray}
    P'(i|k)
    &=&\frac{P(k|i)p(i)}{P(k)} \nonumber\\
    &=&\frac{P(k|i)p(i)}{(1-q)P(k|1)+qP(k|2)}, 
    \label{eq:P'(i|k)}
\end{eqnarray}
where, 
\begin{eqnarray}
    p(1)&=&1-q,\\
    p(2)&=&q,\\
    P(k)&=&(1-q)P(k|1)+qP(k|2).
\end{eqnarray}

Finally, we derive with our expected probability of error $P_{aveerr}$ and its achievable lower bound $P_{aveme}$ as 
\begin{eqnarray}
    P_{aveerr}
    &=&\mathbb{E}_k[P_{err}^{(k)}] 
    =\sum_k P(k)P_{err}^{(k)} \nonumber\\
    &\geq&\sum_k P(k)P_{me}^{(k)}\nonumber\\
    &=&\frac{1}{2}-\sum_kP(k)\frac{\|\hat{\rho}^{(k)}\|_1}{2}\nonumber\\
    &=:&P_{aveme}.
    \label{eq:P_aveerr}
\end{eqnarray}

\section{The average probability of error and conditional improvement via post-selection}
\subsection{The general inequality}\label{subsec:no go proof}
Building on the setup and definitions introduced in the previous section, we demonstrate that the proposed process degrades the discrimination performance, as expected. In fact, it results in a higher error probability than directly discriminating between the original states $|\psi_1\rangle$ and $|\psi_2\rangle$ using the optimal projective measurement that attains the Helstrom bound (i.e. the minimum probability of error). 

In order to prove the proposed process degrades the discrimination capability, we need to prove that,
\begin{eqnarray}
    P_{aveme}\geq P_{me},
    \label{eq:no go ineq}
\end{eqnarray}
which, by plugging the definitions from Eqs. \eqref{eq:me} and \eqref{eq:P_aveerr}), is equivalent to proving, 
\begin{eqnarray}
    \sum_k P(k)\|\hat{\rho}^{(k)}\|_1\leq\|\hat{\rho}\|_1.
    \label{eq: need to proof}
\end{eqnarray}
Since we only consider pure input states, we can use the identity \cite{Barnett:09}
\begin{eqnarray}
 \label{eq:idenity}   \|a|\phi_0\rangle\langle\phi_0|-b|\phi_1\rangle\langle\phi_1|\|_1=\sqrt{1-4ab|\langle\phi_0|\phi_1\rangle|^2},
\end{eqnarray}
which holds for any pure $|\phi_0\rangle$, $|\phi_1\rangle$, and $\forall a,b\in \mathbb{R}$. Use of Eq. \eqref{eq:idenity} and by denoting $|\Psi_i\rangle=\hat{U}|\psi_i,e\rangle$ for simplicity, the left-hand side of inequality \eqref{eq: need to proof} gives, 
\begin{eqnarray}
    &&\sum_k P(k)\|\hat{\rho}^{(k)}\|_1 \nonumber\\
    &=&\sum_k P(k)\sqrt{1-4P'(1|k)P'(2|k)|\langle\Psi_1^{(k)}|\Psi_2^{(k)}\rangle|^2} \nonumber\\
    &=&\sum_k P(k)\sqrt{1-4(1-q)q\frac{|\langle\Psi_1|\hat{I}\otimes \hat{M}_k|\Psi_2\rangle|^2}{P(k)^2}} \nonumber\\
    &\leq&\sqrt{1-4(1-q)q\sum_k P(k)\Big(\frac{|\langle\Psi_1|\hat{I}\otimes \hat{M}_k|\Psi_2\rangle|}{P(k)}\Big)^2} \nonumber\\
    &\leq&\sqrt{1-4(1-q)q\Big(\sum_k P(k)\frac{|\langle\Psi_1|\hat{I}\otimes \hat{M}_k|\Psi_2\rangle|}{P(k)}\Big)^2} \nonumber\\
    &=&\sqrt{1-4(1-q)q\Big(\sum_k |\langle\Psi_1|\hat{I}\otimes \hat{M}_k|\Psi_2\rangle|\Big)^2} \nonumber\\
    &\leq&\sqrt{1-4(1-q)q\Big(|\sum_k \langle\Psi_1|\hat{I}\otimes \hat{M}_k|\Psi_2\rangle|\Big)^2} \nonumber\\
    &=&\sqrt{1-4(1-q)q|\langle\Psi_1|\Psi_2\rangle|^2} \nonumber\\
    &=&\sqrt{1-4(1-q)q|\langle\psi_1,e|\hat{U}^{\dag}\hat{U}|\psi_2,e\rangle|^2} \nonumber\\
    &=&\sqrt{1-4(1-q)q|\langle\psi_1|\psi_2\rangle|^2} \nonumber\\
    &=&\|\hat{\rho}\|_1. 
    \label{eq:proof of no go}
\end{eqnarray}
The first inequality follows from the concavity of the function $f(x) = \sqrt{x}$ for $x\geq0$, where in our case, $\sqrt{x}$ defines a norm therefore is non-negative. Consequently, the inequality is a result of Jensen's inequality. The second inequality similarly follows from the concavity of $f(x) = -x^2$, the application of Jensen's inequality, and the monotonic increasing property of $\sqrt{x}$. The third inequality is a direct consequence of the triangle inequality. Moreover, from the above proof we can see that Eq. (\ref{eq: need to proof}) achieves equality only when
$\frac{\langle\Psi_1|\hat{I}\otimes\hat{M}_k|\Psi_2\rangle}{P(k)}\equiv const,\ \forall k.$ 

While the above analysis establishes that post-selection cannot improve the discrimination performance \emph{on average}, it does not exclude the possibility of conditional improvements within specific post-measurement outcomes. In particular, certain post-selected states may exhibit error probabilities that are strictly lower than the Helstrom bound, even though the ensemble-averaged error remains bounded by it. To illustrate this conditional advantage, we consider the next example.

\subsection{Multimode environment and measurements}\label{subsec:multimode}
We note that the above derivation is not restricted to a single-mode environment.
All steps remain valid for an arbitrary environment Hilbert space $\mathcal{H}_E$,
including multimode settings $\mathcal{H}_E=\bigotimes_{j=1}^m \mathcal{H}_{E_j}$. 
In particular, the environment input may be chosen as a multimode product state
$|e\rangle=\bigotimes_{j=1}^m |e_j\rangle$ (more generally, any normalized pure state in $\mathcal{H}_E$), and the POVM $\{\hat M_k\}$ may act jointly on the full $\mathcal{H}_E$ (with local or product POVMs as special cases). 
The proof of \eqref{eq:proof of no go} only uses the completeness $\sum_k \hat M_k=\hat I_E$ and the normalization $\langle e|e\rangle=1$, hence the inequalities \eqref{eq: need to proof} and \eqref{eq:proof of no go} are valid for the multimode scenario. We depict such setup in Fig. \ref{fig:general-model}.

\subsection{A setup beating the Helstrom Bound: Perfect PNR detector}\label{subsec: Example}

\begin{figure}[ht]
    \centering
    \includegraphics[width=0.8\textwidth]{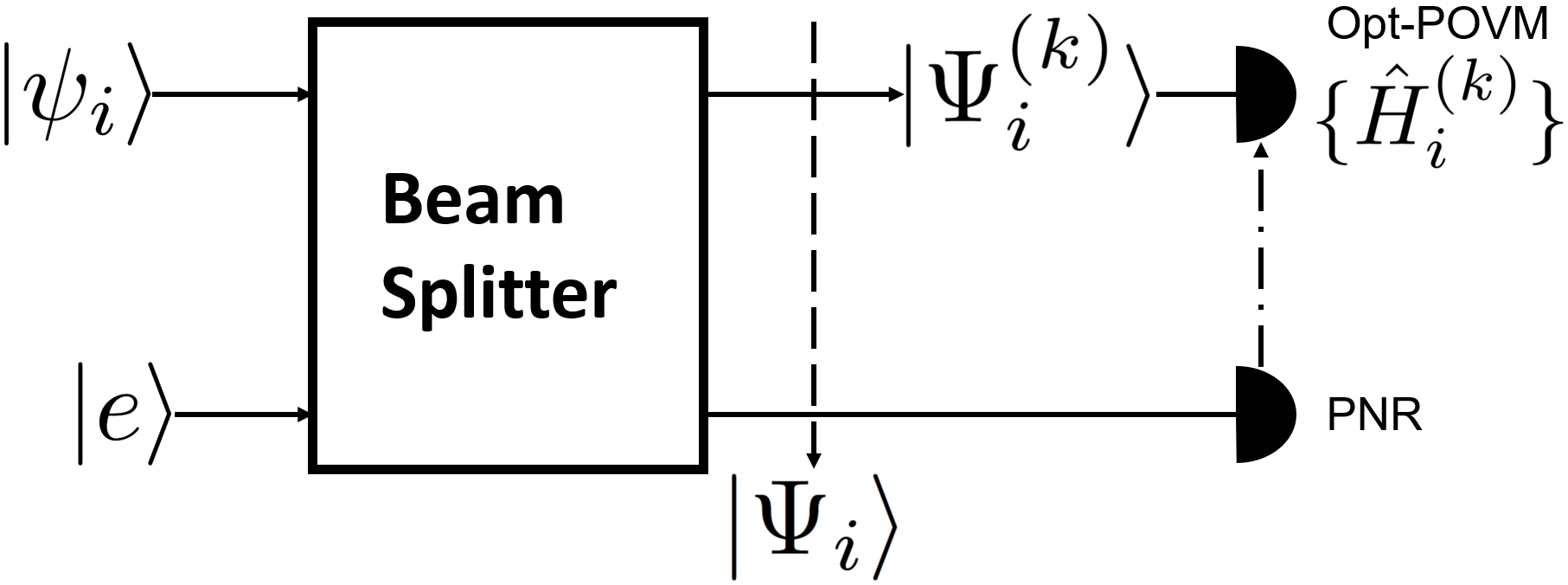}
    \caption{The input state $|\psi_i\rangle$ is an equal superposition of $|0\rangle$ and $|1\rangle$ Fock states. The environment state is fixed to Fock state $|e\rangle=|2\rangle$ in the first example and coherent state $|e\rangle=|\alpha\rangle$ for some $\alpha\in\R$ in the second example, the unitary is chosen as a beam-splitter with transmissivity $\eta$, and the partial measurement is projection on Fock states, i.e., $\{\hat{M}_k=|k\rangle\langle k|\}$. Helstrom measurement $\{\hat{H}_i^{(k)}\}$ conditioned on PNR result $k$ is applied to the discrimination task.}
    \label{fig:exmp-model}
\end{figure}

We consider the following input states, $|\psi_1\rangle=|0\rangle$, $|\psi_2\rangle=\cos\theta|0\rangle+\sin\theta|1\rangle$, where we set $\theta=\frac{\pi}{4}$ for all the evaluation presented in this paper. The environment state in two examples are Fock state $|e\rangle=|2\rangle$ and coherent state $|e\rangle=|\alpha\rangle$ for some $\alpha\in\R$ correspondingly. The unitary transformation $\hat{U}$ is chosen to be a beam splitter with transmissivity $\eta$. This $\hat{U}$ is followed by a photon-number-resolving (PNR) measurement on the second mode. Finally, Helstrom measurement $\{\hat{H}_i^{(k)}\}$ corresponding to the PNR result $k$ is applied to discriminate the remaining state in the first mode to achieve the (averaged) minimum probability of error. The setup is shown in Fig. \ref{fig:exmp-model}. 

\begin{figure}[ht]
  \centering

  \begin{subfigure}[t]{0.45\linewidth}
    \centering
    \includegraphics[width=\linewidth]{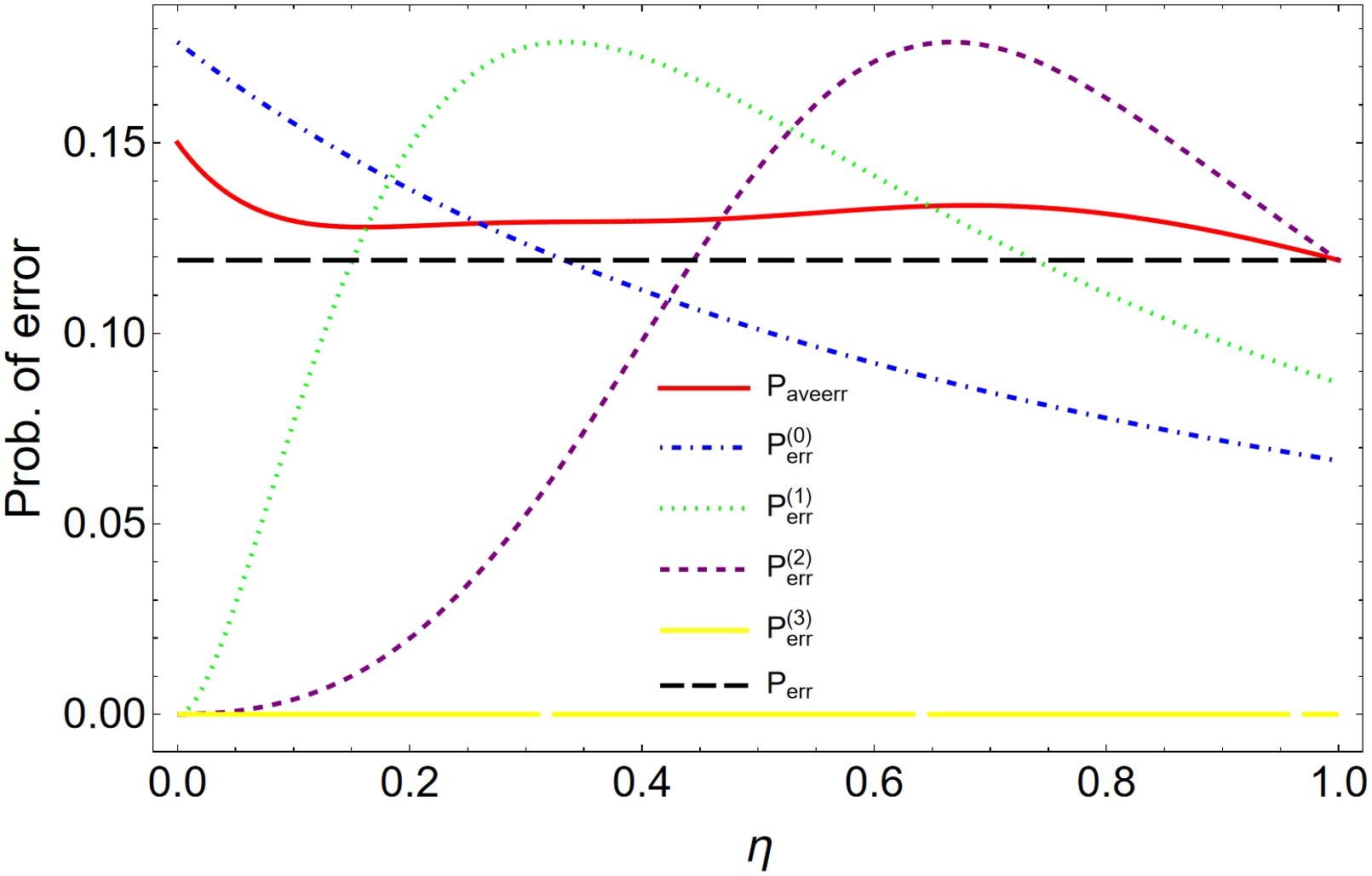}
    \caption{Probability of errors as functions of the transmissivity $\eta$. There exist values of $\eta$ for which post-selection lowers the probability of error compared to $P_{err}$. For example, detection of $k=3$ photons can only come from input state $|\psi_2\rangle$, i.e., the corresponding probability of error is equal to $0$. However, the average performance of state engineering gives a higher than $P_{err}$ probability of error.}
    \label{fig: error-eta}
  \end{subfigure}\qquad
  \begin{subfigure}[t]{0.45\linewidth}
    \centering
    \includegraphics[width=\linewidth]{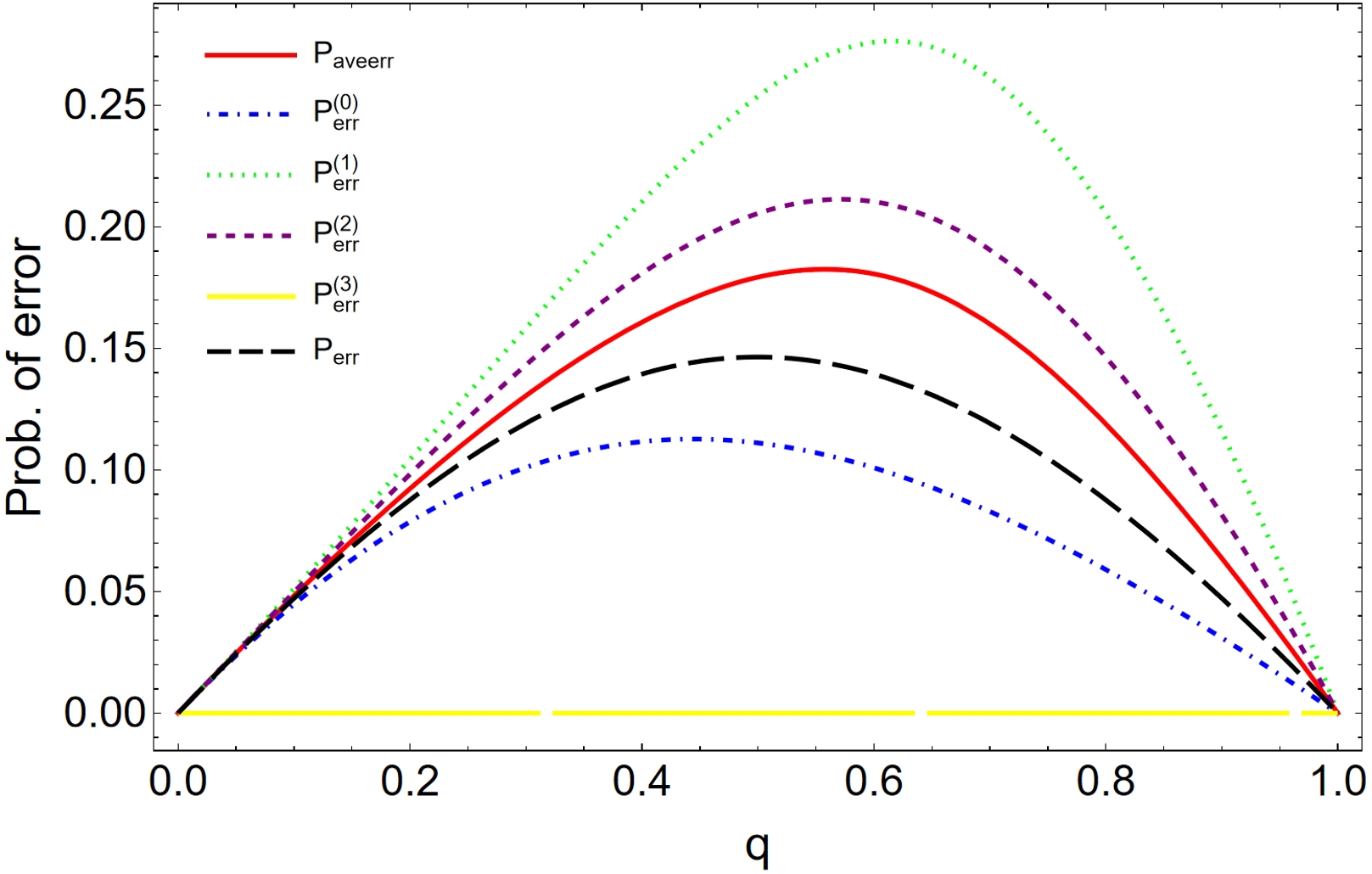}
    \caption{Probability of errors as functions of the prior probability $q$. Even though there exist values of $q$ for which post-selection lowers the probability of error (compared to $P_{err}$), the average performance of state engineering gives a higher than $P_{err}$ probability of error.}
    \label{fig: error-q}
  \end{subfigure}

  \caption{Probability of errors for Example 1. }
  \label{fig:error-ex1}
\end{figure}

\textbf{Example 1:} We use the environment state $|e\rangle=|2\rangle$ in Fig. \ref{fig: error-eta} and Fig. \ref{fig: error-q}. In Fig. \ref{fig: error-eta} we fix the prior probability at $q=0.3$ and plot the probability of errors as functions of the transmissivity $\eta$. In Fig. \ref{fig: error-q}, we fix the transmissivity at $\eta=0.5$ and plot the probability of errors as functions of the prior probability $q$. In both figures, we also plot $P_{err}^{(k)}$ for all PNR outcomes $k\in\{0,1,2,3\}$ and the minimum probability of error $P_{err}$ for the original states $|\Psi_1\rangle$ and $|\Psi_2\rangle$. 

\begin{figure}[ht]
  \centering

  \begin{subfigure}[t]{0.48\textwidth}
    \centering
    \includegraphics[width=\linewidth]{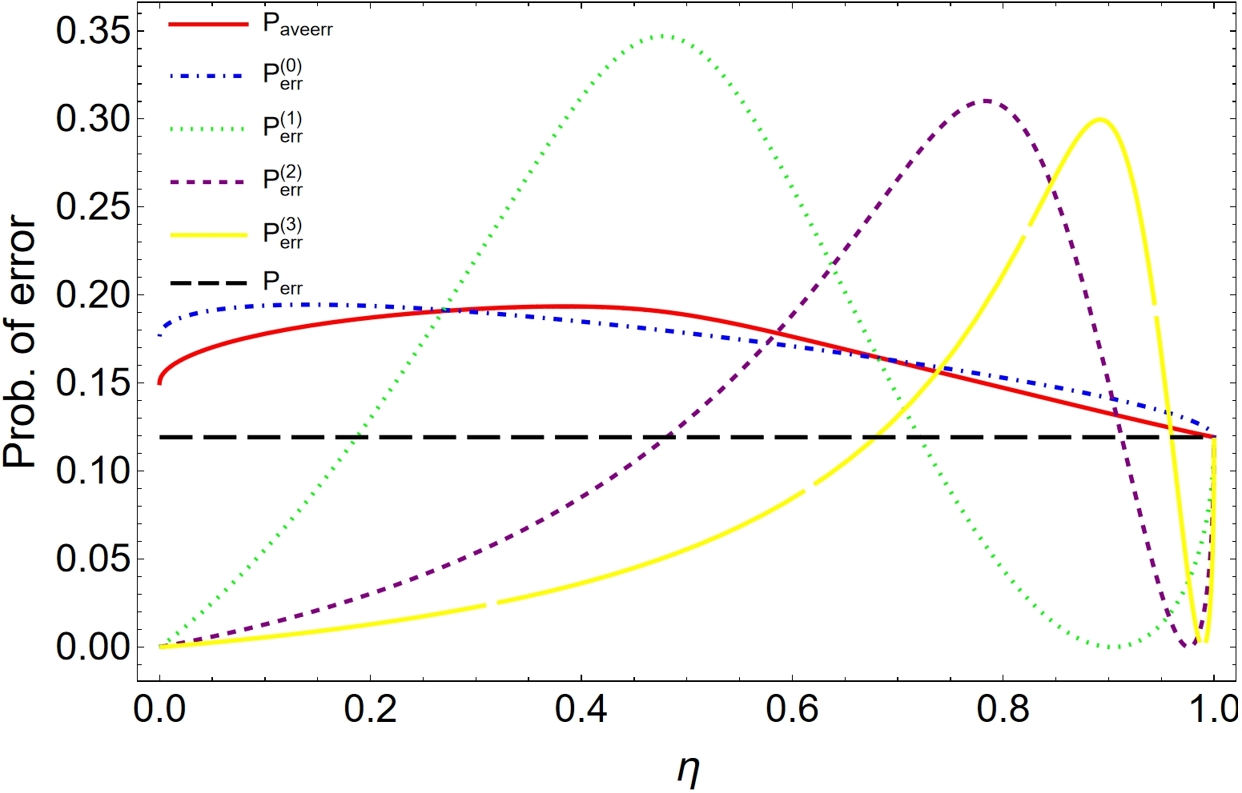}
    \caption{$\alpha=0.3$}
    \label{fig: alpha 0.3}
  \end{subfigure}\hfill
  \begin{subfigure}[t]{0.48\textwidth}
    \centering
    \includegraphics[width=\linewidth]{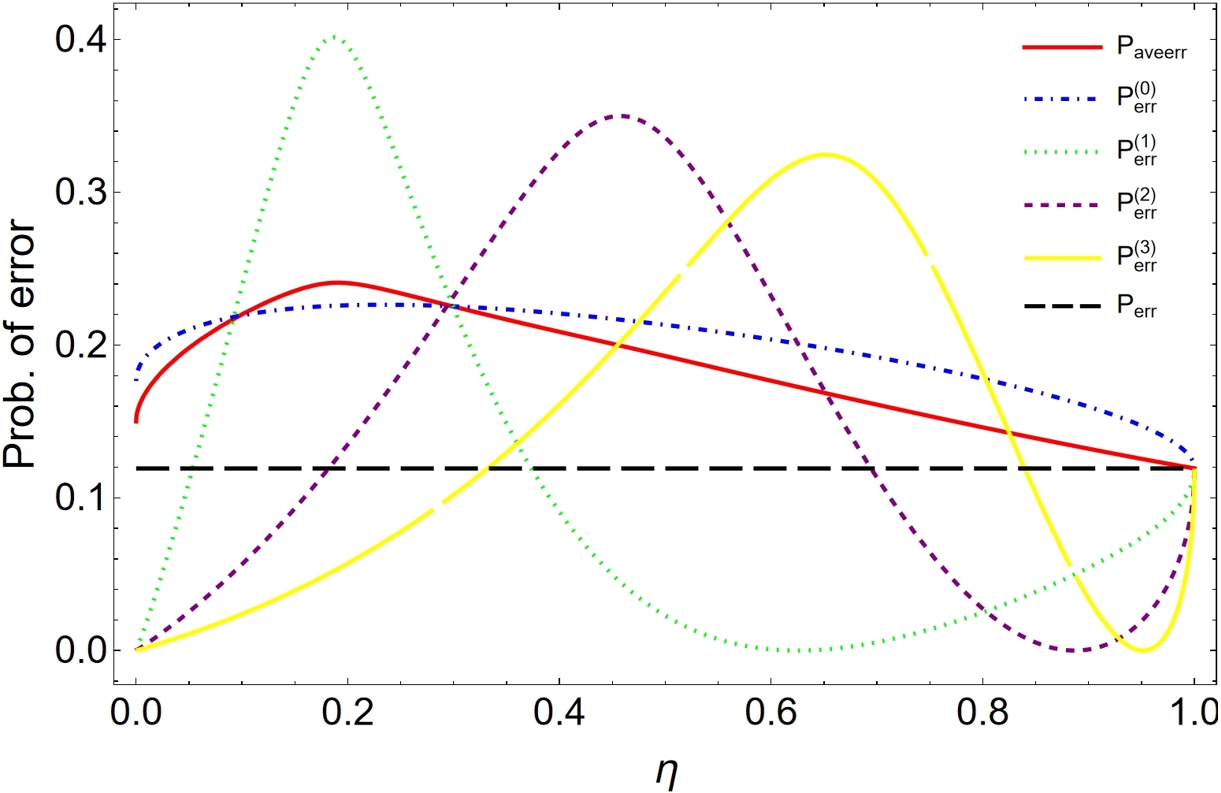}
    \caption{$\alpha=0.6$}
    \label{fig: alpha 0.6}
  \end{subfigure}

  \vspace{0.8em}

  \begin{subfigure}[t]{0.48\textwidth}
    \centering
    \includegraphics[width=\linewidth]{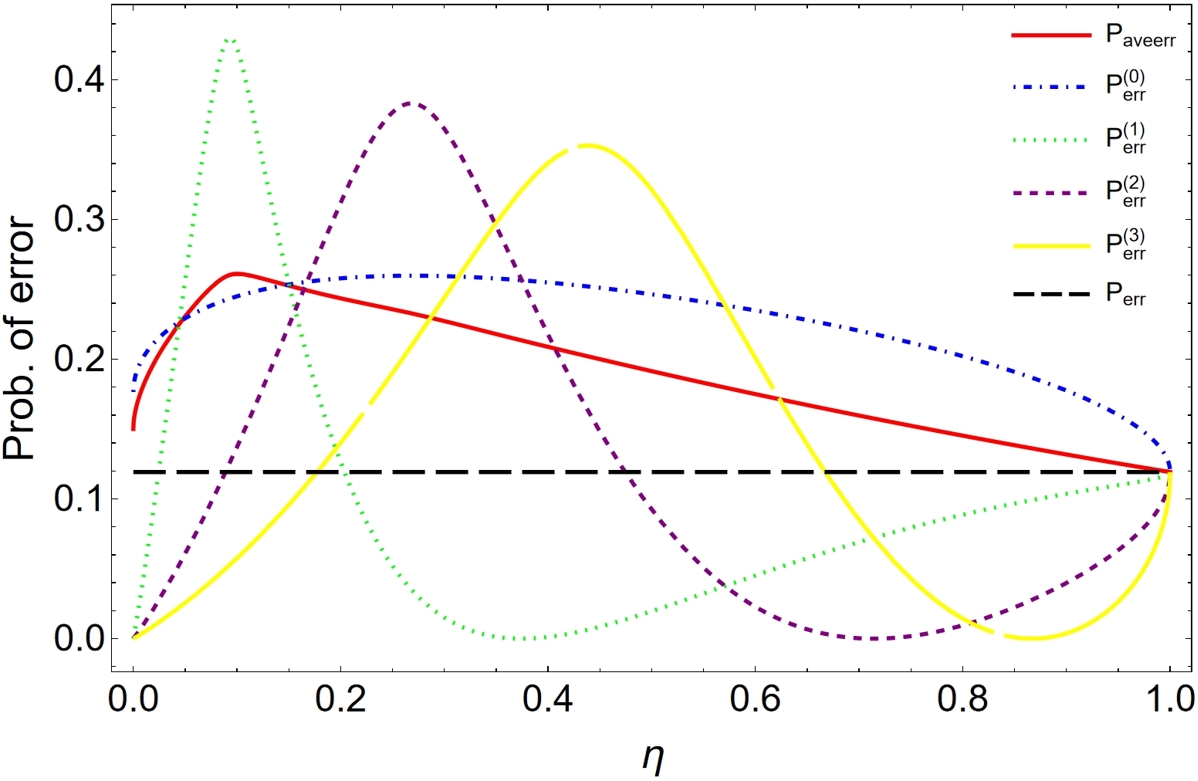}
    \caption{$\alpha=0.9$}
    \label{fig: alpha 0.9}
  \end{subfigure}\hfill
  \begin{subfigure}[t]{0.48\textwidth}
    \centering
    \includegraphics[width=\linewidth]{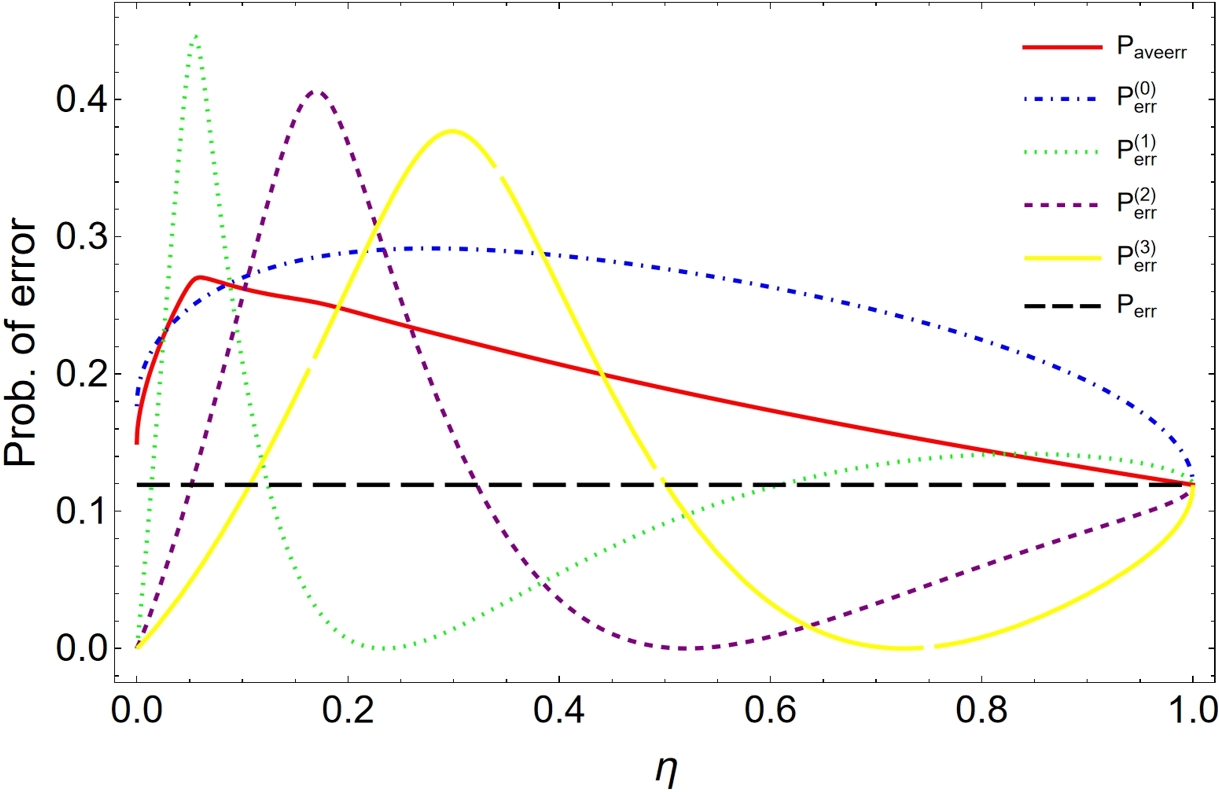}
    \caption{$\alpha=1.2$}
    \label{fig: alpha 1.2}
  \end{subfigure}

  \caption{Probability of errors as functions of the transmissivity $\eta$ with $|e\rangle=|\alpha\rangle$. }
  \label{fig: coherent error-eta}
\end{figure}

\textbf{Example 2:} We use the environment state $|e\rangle=|\alpha\rangle$ with $\alpha=0.3,0.6,0.9,1.2$ in Fig. \ref{fig: coherent error-eta}. In each sub-figure of Fig. \ref{fig: coherent error-eta}, we fix the prior probability at $q=0.3$ and plot the probability of errors as functions of the transmissivity $\eta$ for different $\alpha$. We only plot $P_{err}^{(k)}$ for PNR outcomes $k\in\{0,1,2,3\}$ and the minimum probability of error $P_{err}$ for the original states $|\Psi_1\rangle$ and $|\Psi_2\rangle$. 

We see that in both examples although the average probability of error, $P_{aveerr}$, is always higher than $P_{err}$, the probability of error $P_{err}^{(k)}$ in discriminating the conditional state for each PNR result $k$ is sometimes strictly lower than $P_{err}$. This demonstrates that while post-selection cannot improve the discrimination performance on average, it can enable strictly enhanced discrimination performance in certain post-selected branches. 
The calculation details of this subsection can be found in Appendix A for the first example and Appendix C for the second example. 

This observation indicates that post-selection does not merely filter measurement outcomes but effectively reshapes the statistical structure of the discrimination task: while the average performance remains limited by the Helstrom bound, one can achieve higher distinguishability in a heralded and conditional fashion. If one were to retain only these favorable branches and discard the others, the overall error probability could be further reduced at the expense of a smaller success probability—an operational trade-off reminiscent of, but distinct from, unambiguous state discrimination (USD), where inconclusive outcomes are introduced to guarantee error-free conclusive results (e.g. only retain the branch of $0$ error shown in Fig. \ref{fig:error-ex1}). From this viewpoint, USD can be seen as an extreme limit of our post-selection framework, reached when the conditional error tends to zero—as in the first example, the case of PNR outcome $k=3$—while the success probability decreases. 

\subsection{A setup beating the Helstrom Bound: Lossy PNR detector}\label{subsec: lossy}

\begin{figure}[H]
    \centering
    \includegraphics[width=0.8\textwidth]{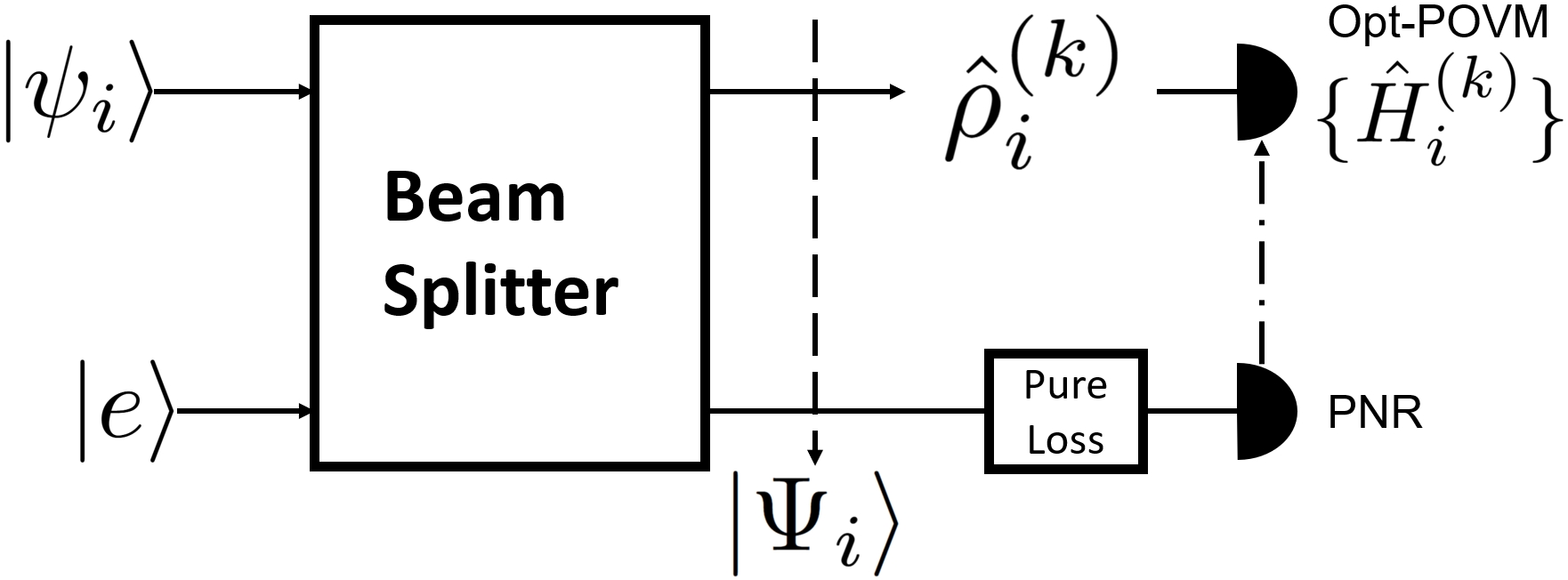}
    \caption{The input state $|\psi_i\rangle$ is an equal superposition of $|0\rangle$ and $|1\rangle$ Fock states. The environment state is Fock state $|e\rangle=|2\rangle$, the unitary is chosen as a beam-splitter with transmissivity $\eta$, and before the PNR measurement there is a pure loss channel. Helstrom measurement $\{\hat{H}_i^{(k)}\}$ conditioned on PNR result $k$ is applied to the discrimination task. The remain state in first mode in this example will be mixed. }
    \label{fig:exmp-model-lossy}
\end{figure}

In this subsection, we consider the following input states, $|\psi_1\rangle=|0\rangle$, $|\psi_2\rangle=\cos\theta|0\rangle+\sin\theta|1\rangle$, where we set $\theta=\frac{\pi}{4}$ for all the evaluation presented in this paper. The environment state is Fock state $|e\rangle=|2\rangle$. The unitary transformation $\hat{U}$ is chosen to be a beam splitter with transmissivity $\eta$. This $\hat{U}$ is followed by a Lossy PNR measurement on the second mode with pure loss channel of transmissivity $\tau$. Finally, Helstrom measurement $\{\hat{H}_i^{(k)}\}$ corresponding to the PNR result $k$ is applied to discriminate the remaining state in the first mode to achieve the (averaged) minimum probability of error. The setup is shown in Fig. \ref{fig:exmp-model-lossy}. 

\begin{figure}[ht]
  \centering

  \begin{subfigure}[t]{0.48\textwidth}
    \centering
    \includegraphics[width=\linewidth]{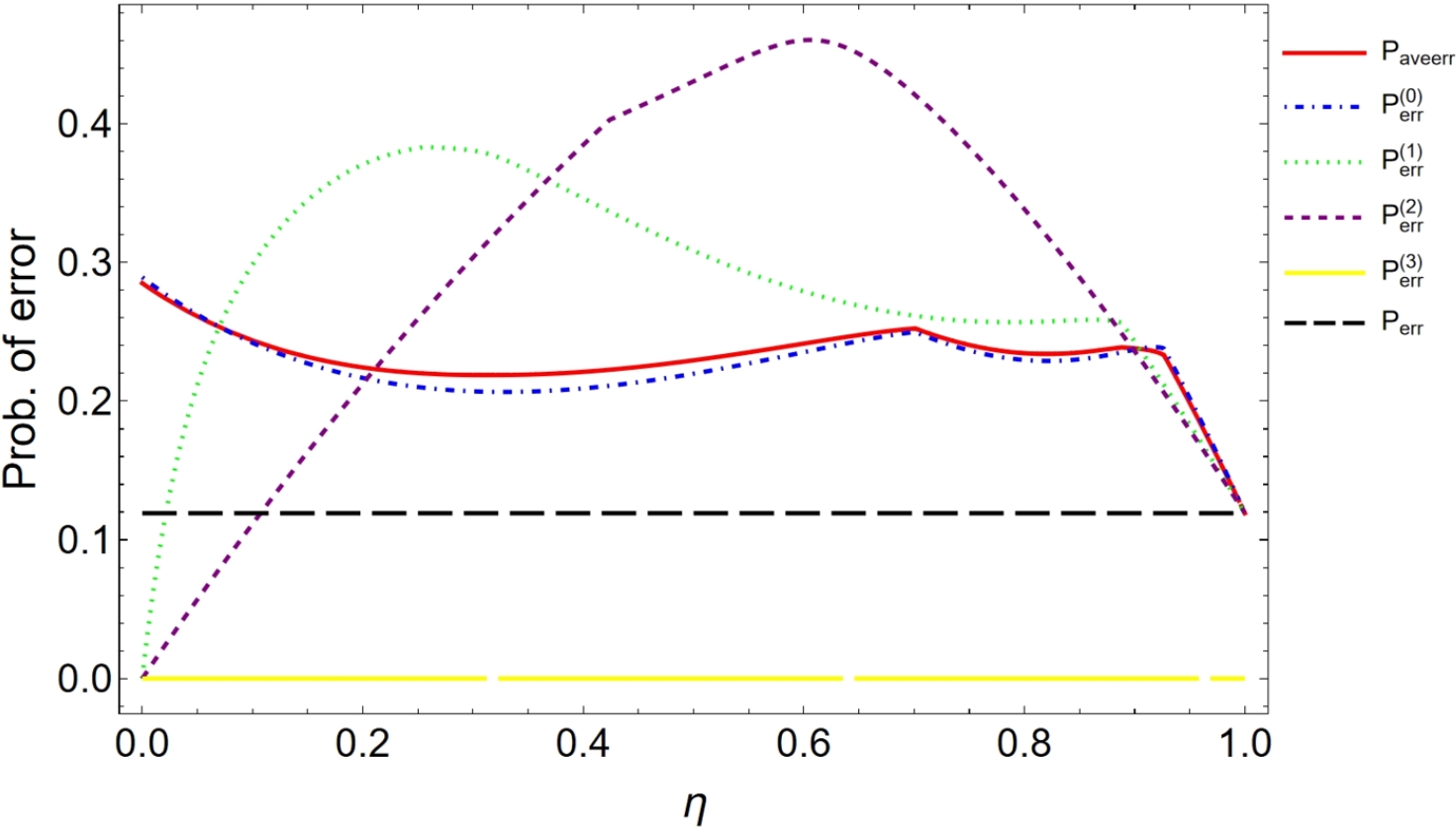}
    \caption{$\tau=0.1$}
    \label{fig: tau 0.1}
  \end{subfigure}\hfill
  \begin{subfigure}[t]{0.48\textwidth}
    \centering
    \includegraphics[width=\linewidth]{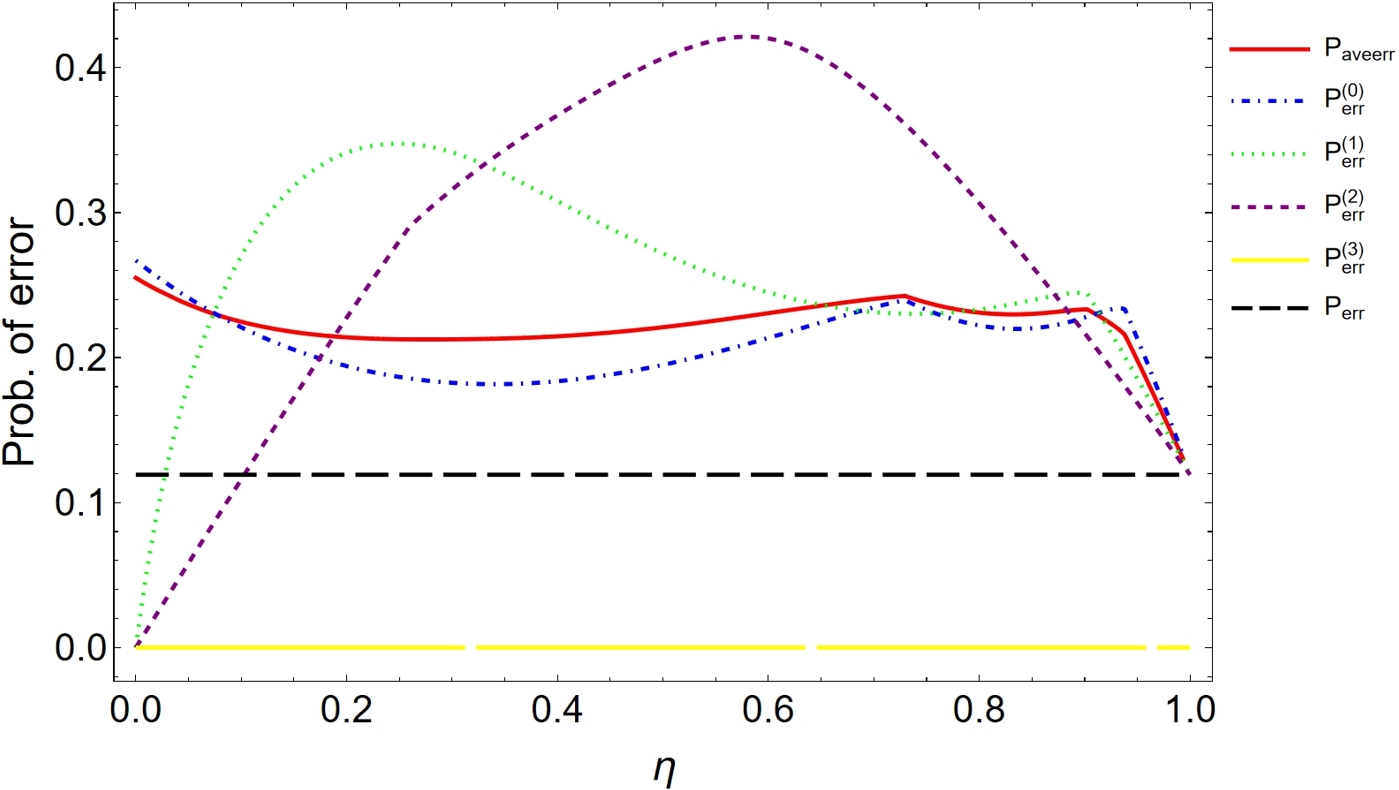}
    \caption{$\tau=0.3$}
    \label{fig: tau 0.3}
  \end{subfigure}

  \vspace{0.8em}

  \begin{subfigure}[t]{0.48\textwidth}
    \centering
    \includegraphics[width=\linewidth]{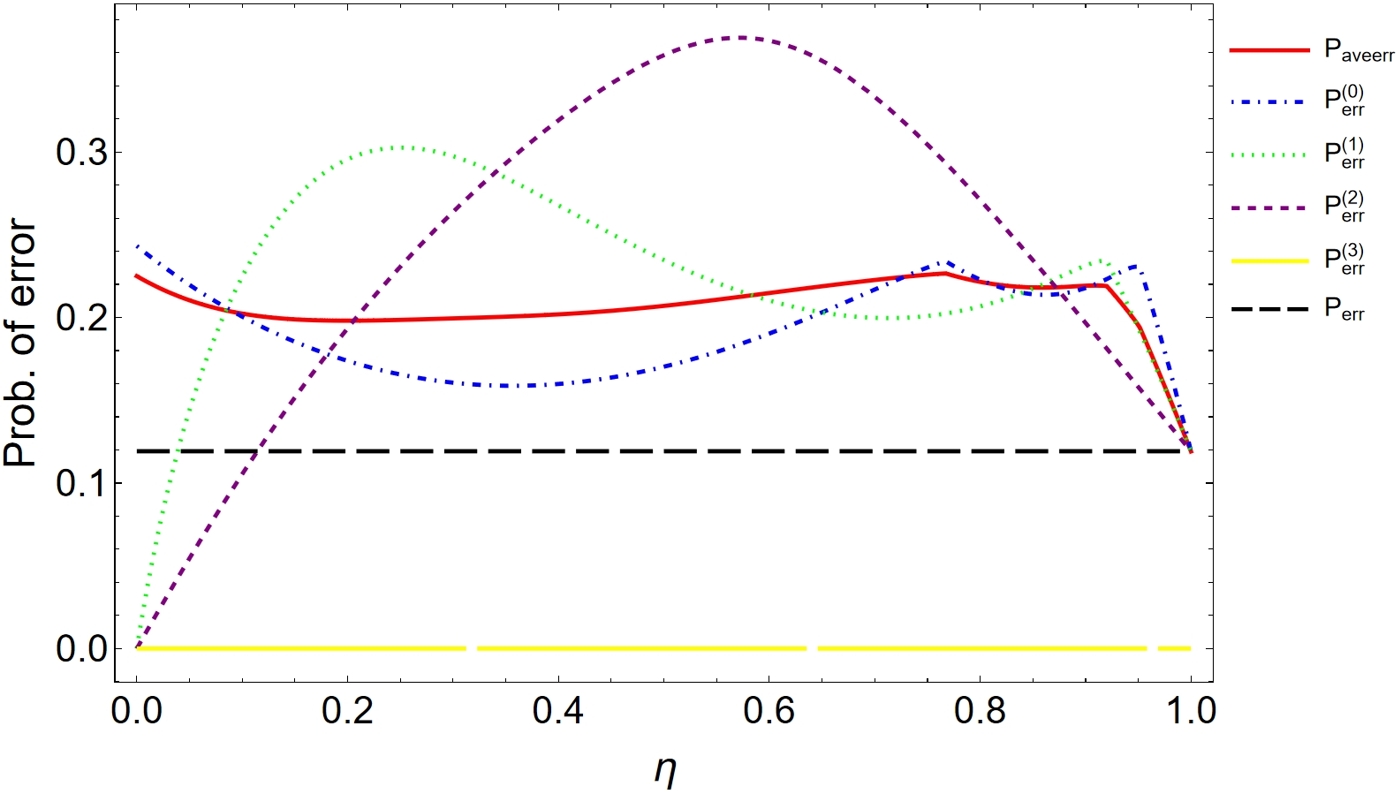}
    \caption{$\tau=0.5$}
    \label{fig: tau 0.5}
  \end{subfigure}\hfill
  \begin{subfigure}[t]{0.48\textwidth}
    \centering
    \includegraphics[width=\linewidth]{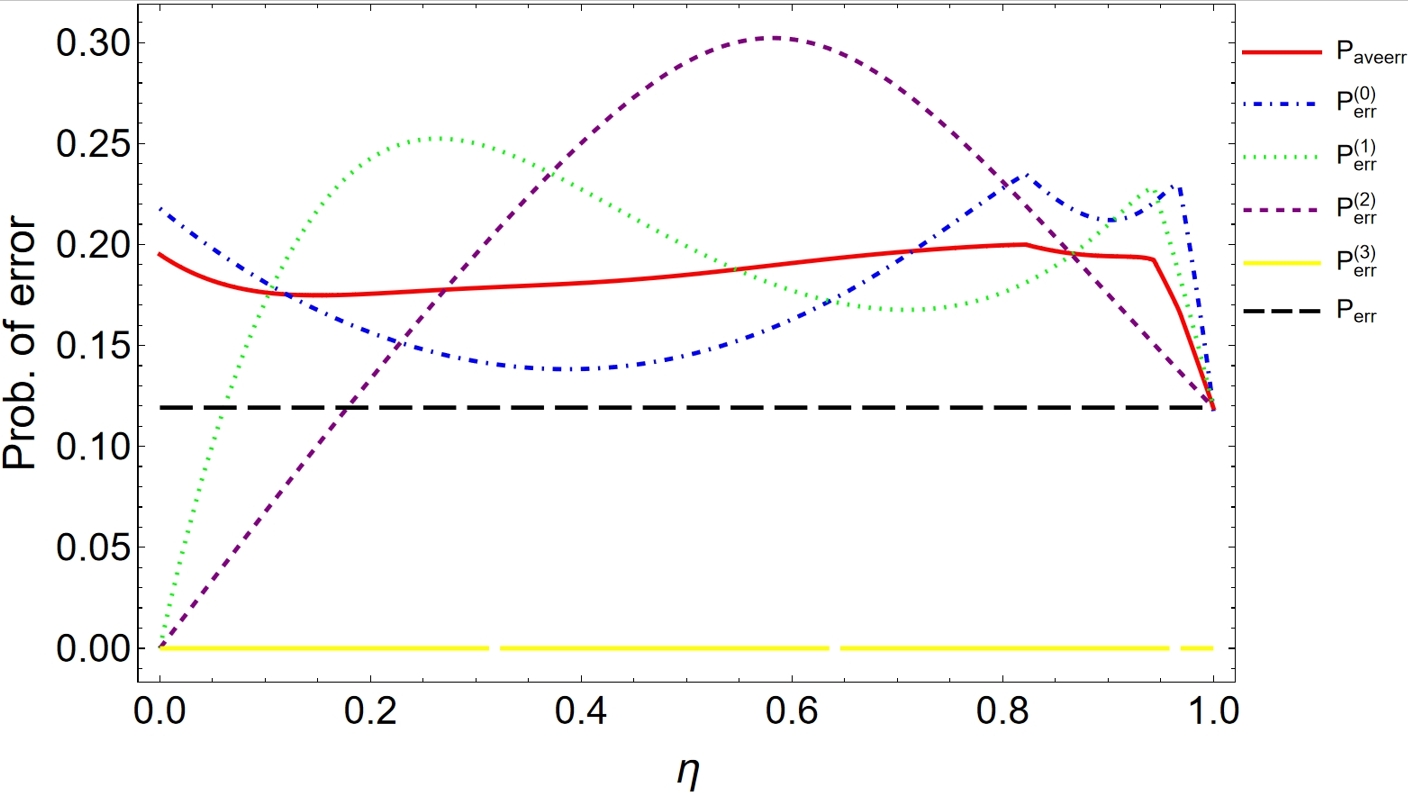}
    \caption{$\tau=0.7$}
    \label{fig: tau 0.7}
  \end{subfigure}

  \caption{Probability of errors as functions of the transmissivity $\eta$ with $|e\rangle=|2\rangle$. }
  \label{fig: lossy error-eta}
\end{figure}

In each sub-figure of Fig. \ref{fig: lossy error-eta}, we fix the prior probability at $q=0.3$ and plot the probability of errors as functions of the transmissivity $\eta$ for different $\tau$. It is shown in all sub-figures that while the average probability of error, $P_{aveerr}$, is always higher than $P_{err}$, the probability of error $P_{err}^{(k)}$ in discriminating the conditional state for each PNR result $k$ is sometimes strictly lower than $P_{err}$, which demonstrates the same behavior as in perfect PNR setting. The calculation details and the description of the pure-loss channel of this subsection can be found in Appendix B. 

This result indicates that our main results in this paper may also be generalized to lossy cases and might have a chance to be proven for mixed states. 

\section{Conclusion}\label{sec:conclusion}

In our setting, where the arbitrary unitary, the arbitrary pure state of the environment, the arbitrary partial measurement, and the classical information transmitted from the lower output mode to the upper output mode are all allowed to inform what measurement should be performed on the conditional state, we have shown the details behind that the minimality of the Helstrom bound, i.e., the optimal performance is based on averaging conditional outcomes. 

Nevertheless, our analysis also uncovers a complementary aspect of this limitation. Although the overall, expectation-based discrimination performance remains bounded by the Helstrom bound, one can find error probabilities strictly lower than the Helstrom bound itself. This conditional enhancement arises because the measurement on the lower output mode generates classical information that determines the subsequent measurement on the upper mode, thereby creating a one-way feed-forward structure within a single-round LOCC framework. Such feed-forward conditioning enables a selective and outcome-dependent refinement of the discrimination strategy: while the global protocol cannot outperform a joint measurement on average, specific conditional outcomes can achieve locally optimal performance surpassing what is attainable without post-selection.

This conditional improvement highlights a potential utility of post-selection-based state engineering as a practical route to explore and harness conditional discrimination advantages. From the theory perspective, it opens a new route to explore general formualae for the heralded discrimination below the Helstrom bound, or to explore generic behavior of important classes of states such as Gaussian states.
\begin{figure}[h]
    \centering
    \includegraphics[width=0.6\textwidth]{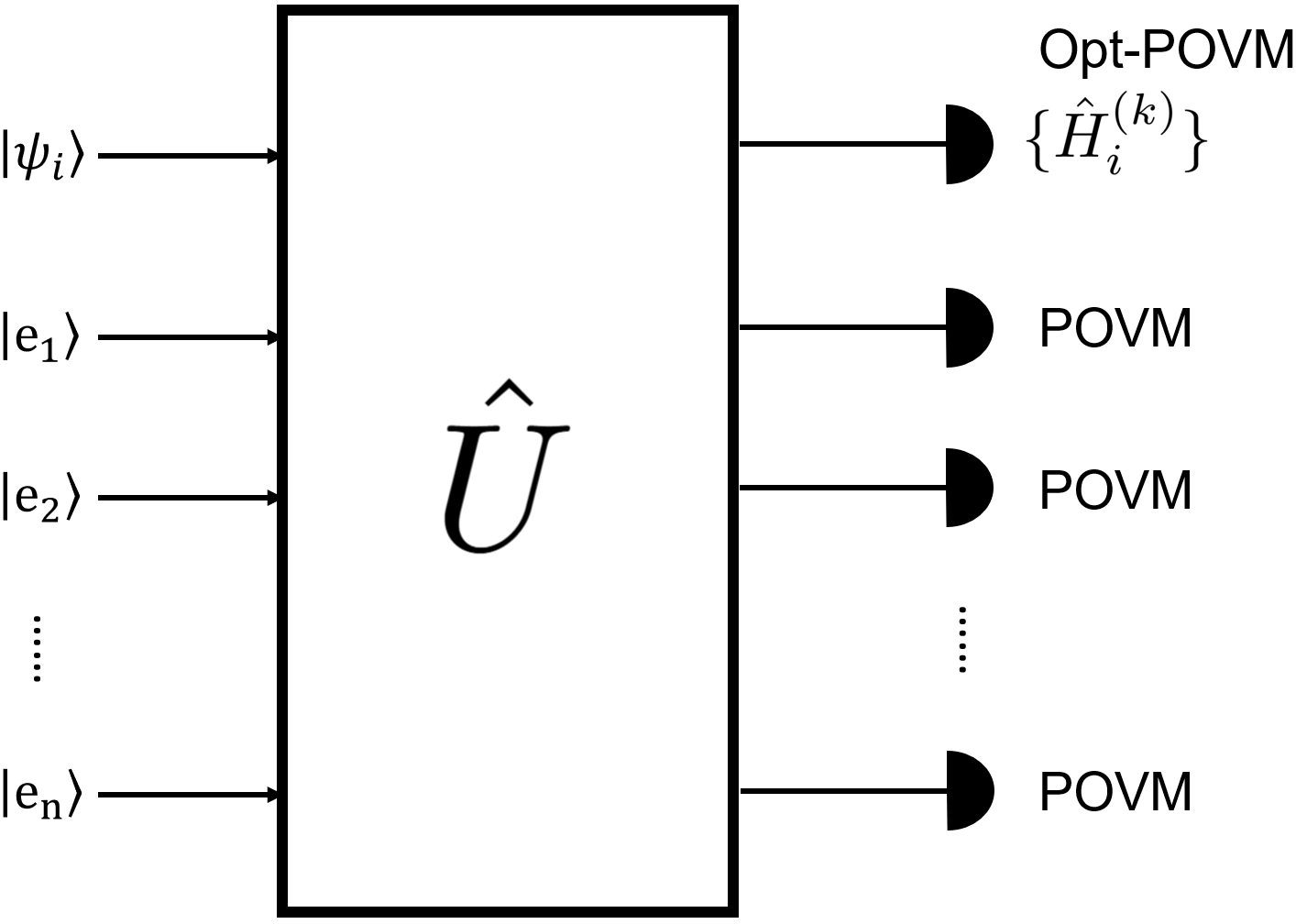}
    \caption{The generic setup: The unitary $\hat{U}$ can in general entangle together the state $|e_1\rangle,\ldots,|e_n\rangle$ or create a joint measurement out of the dis-joint POVMs shown in the setup.}
    \label{fig:general-model}
\end{figure}
As discussed in section \ref{subsec:multimode}, one could start with the most generic scheme (see Fig. \ref{fig:general-model}), for which inequality \eqref{eq:proof of no go} is valid, and post-select on measurement patterns that herald improved discrimination compared to the Helstrom bound. 
\begin{figure}[h]
    \centering
    \includegraphics[width=0.6\textwidth]{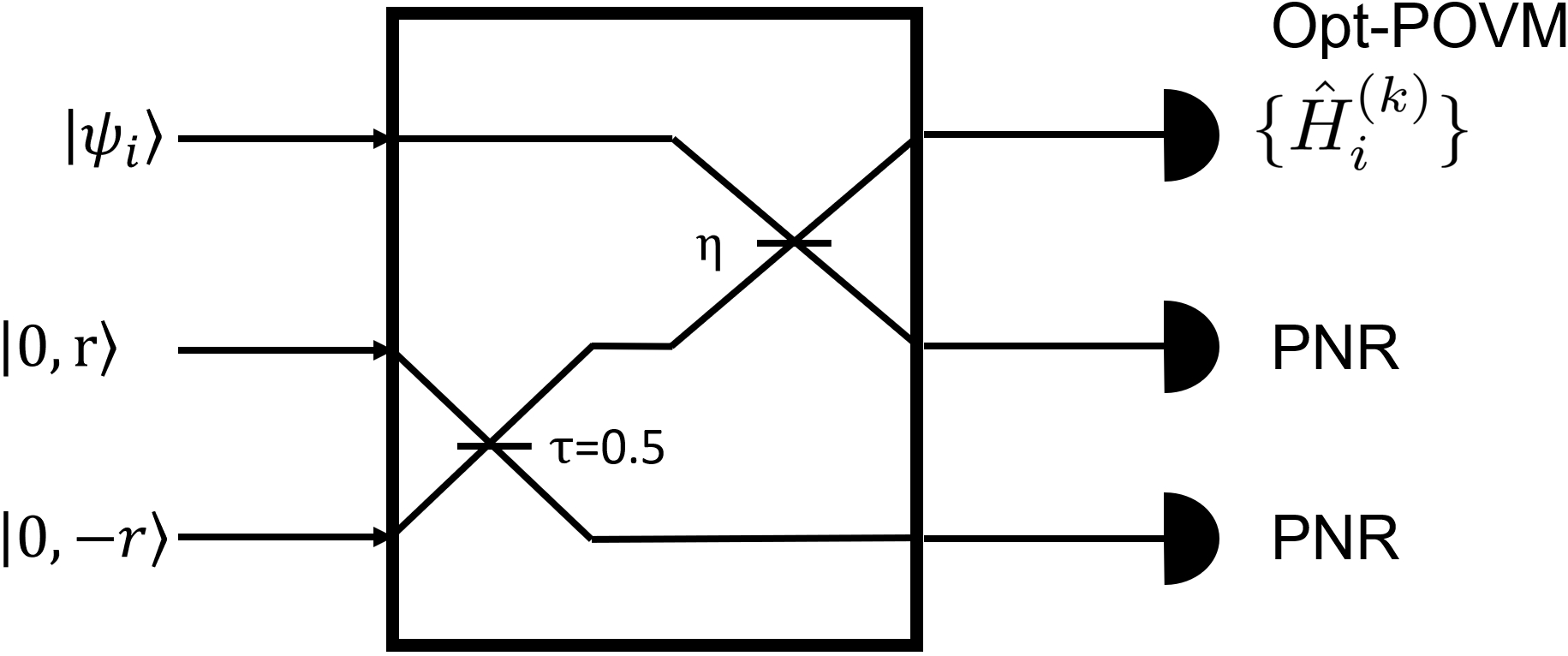}
    \caption{Two squeezed vacuum states with opposite squeezing in the second and third input modes, will generate a two-mode squeezed state. Then upon detecting $k$ photons on the third mode, a Fock $|k\rangle$ will be the state used for our discrimination technique, i.e., it will serve as the environamnt with which the states $|\psi_i\rangle$ will interact under the action of a beam-splitter with transmissivity $\eta$.}
    \label{fig:general-exmp-model}
\end{figure}
We note that the setup in Fig. \ref{fig:exmp-model} can be unfolded to the setup in Fig. \ref{fig:general-exmp-model} which is a subcase of Fig. \ref{fig:general-model}. In Fig. \ref{fig:general-exmp-model}, a two-mode squeezed state (TMSV)is generated and then the lower PNR heralds the Fock state used as the environmental state $|e\rangle$ in section \ref{subsec: Example}, Example 1. Indeed, the TMSV state on the Fock basis assumes the form $|\text{TMSV}\rangle=\sum_{n=0}^\infty c_n(r) |nn\rangle$, therefore detecting $k$ photons on the third mode, will herald $k$ photons on the second mode, which is the lower input of Fig. \ref{fig:exmp-model}. We leave for future works, more general cases based on the setup of Fig. \ref{fig:general-model}.


\funding{National Science Foundation, FET, Award No. 2122337.}

\data{No new data were created or analysed in this study.}

\section*{Appendix A. Derivations of the first example ($|e\rangle=|2\rangle$) in \ref{subsec: Example}}\label{appdxA}
\renewcommand{\thesubsection}{\arabic{subsection}}
\def\theequation{A\arabic{equation}}
\setcounter{equation}{0}
\renewcommand{\thefigure}{A\arabic{figure}}    
\setcounter{figure}{0}
In this Appendix, we show the calculation details of the example shown in Section \ref{subsec: Example}. In said example we work with, $|\psi_1\rangle=|0\rangle$, $|\psi_2\rangle=\cos\theta|0\rangle+\sin\theta|1\rangle$, and the environment state is $|e\rangle=|2\rangle$, where $|0\rangle, |1\rangle, |2\rangle$ are Fock states. Now, we calculate the expected probability of error step by step. 

After passing through $\hat{U}$, we have the outcome two-mode states as 
\begin{eqnarray}
    |\Psi_1\rangle&=&(1-\eta)|20\rangle+\sqrt{2\eta(1-\eta)}|11\rangle+\eta|02\rangle,\\
    |\Psi_2\rangle&=& \cos\theta \Big( (1-\eta)|20\rangle+\sqrt{2\eta(1-\eta)}|11\rangle+\eta|02\rangle \Big) \nonumber\\
    &&+ \sin\theta \Big( \sqrt{3\eta} (1 - \eta) |30\rangle + ( 2\eta \sqrt{(1 - \eta)}-(1 - \eta)^{3/2} ) |21\rangle \nonumber\\
    &&\quad+ (\eta^{3/2} - 2(1 - \eta)\sqrt{\eta}) |12\rangle - \eta \sqrt{3(1-\eta)} |03\rangle \Big). 
\end{eqnarray}
Then, we calculate the probability of getting measurement outcome $k$ given the input state is $|\psi_1\rangle$ and the corresponding remaining state in the first mode as
\begin{eqnarray}
    P(0|1)=(1-\eta)^2&,&|\Psi_1^{(0)}\rangle=|2\rangle;\\
    P(1|1)=2\eta(1-\eta)&,&|\Psi_1^{(1)}\rangle=|1\rangle;\\
    P(2|1)=\eta^2&,&|\Psi_1^{(2)}\rangle=|0\rangle;\\
    P(3|1)=0&;&
\end{eqnarray}
and similarly for $|\psi_2\rangle$
\begin{eqnarray}
    P(0|2)&=&(1-\eta)^2\cos^2\theta+3\eta(1-\eta)^2\sin^2\theta,\\
    |\Psi_2^{(0)}\rangle&=&\frac{(1-\eta)\cos\theta|2\rangle+\sqrt{3\eta}(1-\eta)\sin\theta|3\rangle}{\sqrt{P(0|2)}};\\
    P(1|2)&=&2\eta(1-\eta)\cos^2\theta+( 2\eta \sqrt{(1 - \eta)}-(1 - \eta)^{3/2} )^2\sin^2\theta,\\
    |\Psi_2^{(1)}\rangle&=&\frac{\sqrt{2\eta(1-\eta)}\cos\theta|1\rangle}{\sqrt{P(1|2)}}+\frac{( 2\eta \sqrt{(1 - \eta)}-(1 - \eta)^{3/2} )\sin\theta|2\rangle}{\sqrt{P(1|2)}};\\
    P(2|2)&=&\eta^2\cos^2\theta+(\eta^{3/2} - 2(1 - \eta)\sqrt{\eta})^2\sin^2\theta,\\
    |\Psi_2^{(2)}\rangle&=&\frac{\eta\cos\theta|0\rangle+(\eta^{3/2} - 2(1 - \eta)\sqrt{\eta})\sin\theta|1\rangle}{\sqrt{P(2|2)}};\\
    P(3|2)&=&3\eta^2(1-\eta)\sin^2\theta,\\
    |\Psi_2^{(3)}\rangle&=&\frac{\eta\sqrt{3(1-\eta)}\sin\theta|0\rangle}{\sqrt{P(3|2)}}.
\end{eqnarray}
Next, we use Bayes' theorem to calculate the post-selected prior probability $P'(i|k)$,
\begin{eqnarray}
    P(0)&=&(1-q)(1-\eta)^2+q((1-\eta)^2\cos^2\theta+3\eta(1-\eta)^2\sin^2\theta),\\
    P'(1|0)&=&\frac{(1-q)(1-\eta)^2}{P(0)},\\
    P'(2|0)&=&\frac{q((1-\eta)^2\cos^2\theta+3\eta(1-\eta)^2\sin^2\theta)}{P(0)};\\
    P(1)&=&(1-q)2\eta(1-\eta)+q(2\eta(1-\eta)\cos^2\theta+( 2\eta \sqrt{(1 - \eta)}-(1 - \eta)^{3/2} )^2\sin^2\theta),\\
    P'(1|1)&=&\frac{(1-q)2\eta(1-\eta)}{P(1)},\\
    P'(2|1)&=&\frac{q(2\eta(1-\eta)\cos^2\theta+( 2\eta \sqrt{(1 - \eta)}-(1 - \eta)^{3/2} )^2\sin^2\theta)}{P(1)};\\
    P(2)&=&(1-q)\eta^2+q(\eta^2\cos^2\theta+(\eta^{3/2} - 2(1 - \eta)\sqrt{\eta})^2\sin^2\theta),\\
    P'(1|2)&=&\frac{(1-q)\eta^2}{P(2)},\\
    P'(2|2)&=&\frac{q(\eta^2\cos^2\theta+(\eta^{3/2} - 2(1 - \eta)\sqrt{\eta})^2\sin^2\theta)}{P(2)};\\
    P(3)&=&3q\eta^2(1-\eta)\sin^2\theta,\\
    P'(1|3)&=&0,\\
    P'(2|3)&=&1.
\end{eqnarray}
In addition, to calculate the trace norm, we need the overlaps below,
\begin{eqnarray}
    \langle\Psi_1^{(0)}|\Psi_2^{(0)}\rangle&=&\frac{(1-\eta)\cos\theta}{\sqrt{(1-\eta)^2\cos^2\theta+3\eta(1-\eta)^2\sin^2\theta}},\\
    \langle\Psi_1^{(1)}|\Psi_2^{(1)}\rangle&=&\frac{\sqrt{2\eta(1-\eta)}\cos\theta}{\sqrt{2\eta(1-\eta)\cos^2\theta+( 2\eta \sqrt{(1 - \eta)}-(1 - \eta)^{3/2} )^2\sin^2\theta}},\\
    \langle\Psi_1^{(2)}|\Psi_2^{(2)}\rangle&=&\frac{\eta\cos\theta}{\sqrt{\eta^2\cos^2\theta+(\eta^{3/2} - 2(1 - \eta)\sqrt{\eta})^2\sin^2\theta}};
\end{eqnarray}

Finally, we have, 

\begin{eqnarray}
    P_{err}&=&\frac{1}{2}-\frac{\|\hat{\rho}\|_1}{2}=\frac{1}{2}-\frac{\sqrt{1-4(1-q)q|\langle\psi_1|\psi_2\rangle|^2}}{2}=\frac{1}{2}-\frac{\sqrt{1-4(1-q)q\cos^2\theta}}{2},\\
    P_{aveerr}&=&\frac{1}{2}-\sum_kP(k)\frac{\|\hat{\rho}^{(k)}\|_1}{2}=\frac{1}{2}-\sum_{k=0}^2P(k)\frac{\sqrt{1-4P'(1|k)P'(2|k)|\langle\Psi_1^{(k)}|\Psi_2^{(k)}\rangle|^2}}{2}. 
\end{eqnarray}

\section*{Appendix B. Derivations of the lossy PNR}\label{appdxB}
\renewcommand{\thesubsection}{\arabic{subsection}}
\def\theequation{A\arabic{equation}}
\setcounter{equation}{0}
\renewcommand{\thefigure}{A\arabic{figure}}    
\setcounter{figure}{0}

Consider the lossy PNR where there exists a pure loss channel with transmissivity $\tau$ right before the PNR we operate on the second mode. The kraus operators $\{A_\ell\}_{0}^{\infty}$ corresponding to this action can be written as 
\begin{align}
    A_{\ell}=\sqrt{\frac{(1-\tau)^{\ell}}{\ell!}}\tau^{\frac{\hat{N}}{2}}\hat{a}^{\ell}.
    \label{eq: kraus for pure loss}
\end{align}
Thus, after the pure loss channel, state $|\Psi_i\rangle$ will become 
\begin{align}
    \hat{\rho}_i=\sum_{\ell=0}^{\infty}A_{\ell}|\Psi_i\rangle\langle\Psi_i|A_{\ell}^{\dag}.
    \label{eq: rho_i}
\end{align}
Then, let $\{|0\rangle,|1\rangle,|2\rangle,|3\rangle\}$ be a basis, we calculate the probability of getting measurement outcome $k$ given the input state is $|\psi_1\rangle$, i.e., the state after losses but before PNR is $\hat{\rho}_1$. The corresponding remaining state in the first mode 
\begin{eqnarray}
    y_1^{(0)}(0)&=&\big(0,0,1-\eta,0\big)^T\\
    y_1^{(0)}(1)&=&\big(0,\sqrt{2\eta(1-\eta)(1-\tau)},0,0\big)^T\\
    y_1^{(0)}(2)&=&\big(\eta(1-\tau),0,0,0\big)^T\\
    P(0|1)&=&\sum_{j=0}^2\|y_1^{(0)}(j)\|_2^2,\quad \hat{\rho}_1^{(0)}=\frac{\sum_{j=0}^2 y_1^{(0)}(j) \big(y_1^{(0)}(j)\big)^T}{P(0|1)};\\
    y_1^{(1)}(0)&=&\big(0,\sqrt{2\eta\tau(1-\eta)},0,0\big)^T\\
    y_1^{(1)}(1)&=&\big(\eta\sqrt{2\tau(1-\tau)},0,0,0\big)^T\\
     P(0|1)&=&\sum_{j=0}^1\|y_1^{(1)}(j)\|_2^2,\quad \hat{\rho}_1^{(1)}=\frac{\sum_{j=0}^1 y_1^{(1)}(j) \big(y_1^{(0)}(j)\big)^T}{P(0|1)};\\
    P(2|1)&=&\eta^2\tau^2,\\
    \hat{\rho}_1^{(2)}&=&\frac{\eta^2\tau^2|0\rangle\langle0|}{P(2|1)};\\
    P(3|1)&=&0;
\end{eqnarray}
and similarly for $|\psi_2\rangle$, we have 
\begin{eqnarray}
    y_2^{(0)}(0)&=&\big(0, 0, (1-\eta)\cos\theta, \sqrt{3\eta}(1-\eta)\sin\theta \big)^T,\\
    y_2^{(0)}(1)&=&\big(0, \sqrt{2\eta(1-\eta)(1-\tau)}\cos\theta, \sqrt{1-\tau}(2\eta\sqrt{1-\eta}-(1-\eta)^{\frac{3}{2}})\sin\theta, 0\big)^T,\\
    y_2^{(0)}(2)&=&\big( \eta(1-\tau)\cos\theta, (1-\tau)(\eta^{\frac{3}{2}}-2(1-\eta)\sqrt{\eta})\sin\theta, 0, 0\big)^T,\\
    y_2^{(0)}(3)&=&\big( -\eta\sqrt{3(1-\tau)^3(1-\eta)}\sin\theta, 0, 0, 0 \big)^T,\\
    P(0|2)&=&\sum_{j=0}^3\|y_2^{(0)}(j)\|_2^2,\quad \hat{\rho}_2^{(0)}=\frac{\sum_{j=0}^3 y_2^{(0)}(j) \big(y_2^{(0)}(j)\big)^T}{P(0|2)};\\
    y_2^{(1)}(0)&=&\big(  0, \sqrt{2\eta\tau(1-\eta)}\cos\theta, \sqrt{\tau}(2\eta\sqrt{1-\eta}-(1-\eta)^{\frac{3}{2}})\sin\theta, 0 \big)^T,\\
    y_2^{(1)}(1)&=&\big( \eta\sqrt{2\tau(1-\tau)}\cos\theta, \sqrt{2\tau(1-\tau)}(\eta^{\frac{3}{2}}-2(1-\eta)\sqrt{\eta})\sin\theta, 0, 0 \big)^T,\\
    y_2^{(1)}(2)&=&\big( -3\eta(1-\tau)\sqrt{\tau(1-\eta)}\sin\theta,0,0,0 \big)^T,\\
    P(1|2)&=&\sum_{j=0}^2\|y_2^{(1)}(j)\|_2^2,\quad \hat{\rho}_2^{(1)}=\frac{\sum_{j=0}^2 y_2^{(1)}(j) \big(y_2^{(1)}(j)\big)^T}{P(1|2)};\\
    y_2^{(2)}(0)&=&\big( \eta\tau\cos\theta, \tau(\eta^{\frac{3}{2}}-2(1-\eta)\sqrt{\eta})\sin\theta, 0, 0 \big)^T,\\
    y_2^{(2)}(1)&=&\big( -3\eta\tau\sqrt{(1-\eta)(1-\tau)}\sin\theta, 0, 0, 0 \big)^T,\\
    P(2|2)&=&\sum_{j=0}^1\|y_2^{(2)}(j)\|_2^2,\quad \hat{\rho}_2^{(2)}=\frac{\sum_{j=0}^1 y_2^{(2)}(j) \big(y_2^{(2)}(j)\big)^T}{P(2|2)};\\
    y_2^{(3)}(0)&=&\big( -\eta\sqrt{3\tau^3(1-\eta)}\sin\theta,0,0,0 \big)^T,\\
    P(3|2)&=&\|y_2^{(3)}(0)\|_2^2,\quad \hat{\rho}_2^{(3)}=\frac{y_2^{(3)}(0) \big(y_2^{(3)}(0)\big)^T}{P(3|2)}.
\end{eqnarray}
Then, we use Bayes' theorem to calculate the post-selected prior probability $P'(i|k)$ and finally get 
\begin{align}
    P_{aveerr}^{\text{noisy}}=\frac{1}{2}-\sum_{k=0}^2 P(k)\frac{\|P'(1|k)\hat{\rho}_1^{(k)}-P'(2|k)\hat{\rho}_2^{(k)}\|_1}{2}.
\end{align}

\section*{Appendix C. Derivations of the second example ($|e\rangle=|\alpha\rangle$) in \ref{subsec: Example}}\label{appdxC}
\renewcommand{\thesubsection}{\arabic{subsection}}
\def\theequation{A\arabic{equation}}
\setcounter{equation}{0}
\renewcommand{\thefigure}{A\arabic{figure}}    
\setcounter{figure}{0}


Consider a beam splitter of transmissivity $\eta\in[0,1]$ acting on two modes $a$ and $b$.
For $|\psi_1\rangle=|0\rangle$, the output of the beam splitter is
\begin{align}
    |\Psi_1\rangle
    =U_{\mathrm{BS}}|0\rangle_a|\alpha\rangle_b
    =|\sqrt{1-\eta}\alpha\rangle_a |\sqrt{\eta}\alpha\rangle_b.
\label{eq:BS-vac-coh}
\end{align}
Importantly, the output remains a product of coherent states, hence no entanglement is created between the two modes in this case.

PNR on the 2nd mode yields outcome $k\in\mathbb{N}$,
the corresponding probability is Poisson with mean $\eta|\alpha|^2$:
\begin{align}
    P(k|1)=e^{-\eta|\alpha|^2}\frac{\bigl(\eta|\alpha|^2\bigr)^k}{k!}.
\label{eq:pk-poisson}
\end{align}
The conditional state is independent of $k$:
\begin{align}
|\Psi_1^{(k)}\rangle
=
\frac{_b\langle k|U_{\mathrm{BS}}(|0\rangle_a|\alpha\rangle_b)}{\sqrt{P(k|1)}}
=
|\sqrt{1-\eta}\alpha\rangle.
\label{eq:norm-cond}
\end{align}

For $|\psi_2\rangle=\cos\theta|0\rangle+\sin\theta|1\rangle$, the output of the beam splitter will be 
\begin{align}
|\Psi_2\rangle
=&\ \cos\theta\,
|\sqrt{1-\eta}\alpha\rangle_a|\sqrt{\eta}\alpha\rangle_b \nonumber\\
&+ \sin\theta\,
U_{\mathrm{BS}}\bigl(|1\rangle_a|\alpha\rangle_b\bigr).
\label{eq:psi2-out-split}
\end{align}
Using the beam-splitter relations
\begin{align}
U_{\mathrm{BS}}\,a^\dagger\,U_{\mathrm{BS}}^\dagger
=\sqrt{\eta}\,a^\dagger-\sqrt{1-\eta}\,b^\dagger,
\qquad
U_{\mathrm{BS}}|0\rangle_a|\alpha\rangle_b
=|\sqrt{1-\eta}\alpha\rangle_a|\sqrt{\eta}\alpha\rangle_b,
\label{eq:bs-relations}
\end{align}
we can write
\begin{align}
U_{\mathrm{BS}}\bigl(|1\rangle_a|\alpha\rangle_b\bigr)
&=U_{\mathrm{BS}}\,a^\dagger|0\rangle_a|\alpha\rangle_b
=\bigl(U_{\mathrm{BS}}a^\dagger U_{\mathrm{BS}}^\dagger\bigr)\,
U_{\mathrm{BS}}|0\rangle_a|\alpha\rangle_b \nonumber\\
&=\Bigl(\sqrt{\eta}\,a^\dagger-\sqrt{1-\eta}\,b^\dagger\Bigr)
|\sqrt{1-\eta}\alpha\rangle_a|\sqrt{\eta}\alpha\rangle_b.
\label{eq:bs-on-1coh}
\end{align}
Therefore,
\begin{align}
|\Psi_2\rangle
=&\ \cos\theta\,
|\sqrt{1-\eta}\alpha\rangle_a|\sqrt{\eta}\alpha\rangle_b \nonumber\\
&+\sin\theta\Bigl(\sqrt{\eta}\,a^\dagger-\sqrt{1-\eta}\,b^\dagger\Bigr)
|\sqrt{1-\eta}\alpha\rangle_a|\sqrt{\eta}\alpha\rangle_b.
\label{eq:psi2-out-compact}
\end{align}

After performing PNR on the 2nd mode, the (unnormalized) conditional state of the 1st mode is
\begin{align}
|\widetilde{\Psi}^{(k)}_2\rangle
&:={}_b\langle k|\Psi_2\rangle \nonumber\\
&=\cos\theta\,
{}_b\langle k|\sqrt{\eta}\alpha\rangle_b\,
|\sqrt{1-\eta}\alpha\rangle_a \nonumber\\
&\quad+\sin\theta\Bigl[
\sqrt{\eta}\,{}_b\langle k|\sqrt{\eta}\alpha\rangle_b\,a^\dagger|\sqrt{1-\eta}\alpha\rangle_a
-\sqrt{1-\eta}\,{}_b\langle k|b^\dagger|\sqrt{\eta}\alpha\rangle_b\,|\sqrt{1-\eta}\alpha\rangle_a
\Bigr].
\label{eq:unnorm-cond-start}
\end{align}
We use the coherent-state overlap
\begin{align}
{}_b\langle k|\gamma\rangle_b
=e^{-|\gamma|^2/2}\frac{\gamma^k}{\sqrt{k!}},
\qquad \gamma:=\sqrt{\eta}\alpha,
\label{eq:coh-overlap}
\end{align}
and the identity ${}_b\langle k|b^\dagger=\sqrt{k}\,{}_b\langle k-1|$, to obtain (for $k\ge 1$)
\begin{align}
|\widetilde{\Psi}^{(k)}_2\rangle
=&\ e^{-|\gamma|^2/2}\Bigg[
\Bigl(\cos\theta\,\frac{\gamma^k}{\sqrt{k!}}
-\sin\theta\,\sqrt{1-\eta}\,\sqrt{k}\,\frac{\gamma^{k-1}}{\sqrt{(k-1)!}}\Bigr)\,
|\sqrt{1-\eta}\alpha\rangle_a \nonumber\\
&\hspace{5.2em}
+\sin\theta\,\sqrt{\eta}\,\frac{\gamma^k}{\sqrt{k!}}\,
a^\dagger|\sqrt{1-\eta}\alpha\rangle_a
\Bigg].
\label{eq:unnorm-cond-correct}
\end{align}
For $k=0$, the second term in the first line above is absent and the same expression holds with the convention
$\sqrt{k}\,\gamma^{k-1}/\sqrt{(k-1)!}=0$.

We then calculate the corresponding probability of obtaining outcome $k$.
For convenience, define
\begin{align}
\beta:=\sqrt{1-\eta}\alpha,\qquad
\gamma:=\sqrt{\eta}\alpha
\qquad(\alpha\in\mathbb{R}),
\label{eq:def-beta-gamma}
\end{align}
and rewrite \eqref{eq:unnorm-cond-correct} as
\begin{align}
|\widetilde{\Psi}^{(k)}_2\rangle
=e^{-\gamma^2/2}\Big(a_k|\beta\rangle+b_k\,a^\dagger|\beta\rangle\Big),
\label{eq:unnorm-cond-ab}
\end{align}
where (for $k\ge 0$)
\begin{align}
a_k
:=\cos\theta\,\frac{\gamma^k}{\sqrt{k!}}
-\sin\theta\,\sqrt{1-\eta}\,\sqrt{k}\,\frac{\gamma^{k-1}}{\sqrt{(k-1)!}},
\qquad
b_k
:=\sin\theta\,\sqrt{\eta}\,\frac{\gamma^k}{\sqrt{k!}}.
\label{eq:akbk}
\end{align}
Thus,
\begin{align}
P(k|2)
:=\langle \widetilde{\Psi}^{(k)}_2|\widetilde{\Psi}^{(k)}_2\rangle
=e^{-\gamma^2}\Big(
a_k^2\langle \beta|\beta\rangle
+b_k^2\langle \beta|aa^\dagger|\beta\rangle
+2a_k b_k\,\langle \beta|a^\dagger|\beta\rangle
\Big).
\label{eq:Pk2-start}
\end{align}
Using $\langle \beta|\beta\rangle=1$, $\langle \beta|a^\dagger|\beta\rangle=\beta$ and
$\langle \beta|aa^\dagger|\beta\rangle=\beta^2+1$ (with $\beta\in\mathbb{R}$), we obtain the closed form
\begin{align}
P(k|2)
=e^{-\gamma^2}\Big(a_k^2+b_k^2(\beta^2+1)+2a_k b_k\,\beta\Big),
\qquad (k\ge 0),
\label{eq:Pk2-closed}
\end{align}
with $a_k$ and $b_k$ given in \eqref{eq:akbk}.

Finally, the normalized post-measurement state in the 1st mode conditioned on outcome $k$ is 
\begin{align}
|\Psi_2^{(k)}\rangle
=\frac{|\widetilde{\Psi}_2^{(k)}\rangle}{\sqrt{P(k|2)}}.
\end{align}

Using $\langle\beta|\beta\rangle=1$ and $\langle\beta|a^\dagger|\beta\rangle=\beta$ (for real $\beta$), we obtain
\begin{align}
\langle \Psi_1^{(k)}|\Psi_2^{(k)}\rangle
&=\frac{\langle \beta|\widetilde{\Psi}_2^{(k)}\rangle}{\sqrt{P(k|2)}}
=\frac{a_k+b_k\beta}{\sqrt{a_k^2+b_k^2(\beta^2+1)+2a_k b_k\,\beta}} \notag\\
&=\frac{a_k+b_k\beta}{\sqrt{(a_k+b_k\beta)^2+b_k^2}},\\
\left|\langle \Psi_1^{(k)}|\Psi_2^{(k)}\rangle\right|^2
&=\frac{(a_k+b_k\beta)^2}{(a_k+b_k\beta)^2+b_k^2}.
\end{align}
For equal priors, the Helstrom error probability conditioned on outcome $k$ is
\begin{align}
P_{\mathrm{err}}^{(k)}
=\frac{1}{2}-\frac{\sqrt{1-4P'(1|k)P'(2|k)\left|\langle \Psi_1^{(k)}|\Psi_2^{(k)}\rangle\right|^2}}{2}, 
\end{align}
where 
\begin{align}
P'(i|k)=\frac{(1-q)P(k|i)}{(1-q)P(k|1)+qP(k|2)}. 
\end{align}

\bibliographystyle{iopart-num}
\bibliography{refs}

\end{document}